\documentclass[3p,times,authoryear]{article} % For LaTeX2e
\usepackage{iclr2021_conference,times}
\usepackage{makecell, multirow}
\usepackage{graphicx}
\usepackage[tableposition=below]{caption}
\usepackage{longtable}
\usepackage{amsmath,amsfonts,bm}

\def\eqref#1{equation~\ref{#1}}
\def\1{\bm{1}}

\DeclareMathAlphabet{\mathsfit}{\encodingdefault}{\sfdefault}{m}{sl}
\SetMathAlphabet{\mathsfit}{bold}{\encodingdefault}{\sfdefault}{bx}{n}

\usepackage{enumitem}
\usepackage{hyperref}
\hypersetup{colorlinks,allcolors=black}

\usepackage{fancyhdr}
\usepackage{array}
\usepackage{url}
\usepackage{tabularx}
\usepackage{adjustbox}
\usepackage{array}
\usepackage{float}
\setcitestyle{round,authoryear,comma}
\usepackage{geometry}

\usepackage{array,ragged2e}
\newcolumntype{M}[1]{>{\Centering\arraybackslash}m{#1}}

\makeatletter
\renewcommand\paragraph{\@startsection{paragraph}{4}{0ex}%
   {1mm}%
   {1mm}%
   {\normalfont\normalsize\bfseries}}
\newcommand\mymaketitle{%
  \begin{titlepage}
    \null\vfil\vskip 40\p@
    \begin{center}
      {\@title}
      \vskip 2.5em
      {\large \lineskip .75em \@author \par}
      \vskip 1.5em
      {\large \@date \par}
    \end{center}\par
    \@thanks
    \vfil\null
  \end{titlepage}
}
\makeatother

\usepackage{fancyhdr}

\title{A Multi-Modal Unsupervised Machine Learning Approach for Biomedical Signal Processing in CPR}
\author{Saidul Islam$^1$, Jamal Bentahar$^{2,1,*}$, Robin Cohen$^3$, Gaith Rjoub$^{4,1}$ \\
$^1$Concordia Institute for Information Systems Engineering, 
Concordia University, Montreal, Canada \\
$^2$Department of Computer Science,
Khalifa University, Abu Dhabi, UAE\\
$^3$Cheriton School of Computer Science, University of Waterloo, Waterloo, Canada\\
$^4$Faculty of Information Technology, Aqaba University of Technology, Aqaba, Jordan\\
\\\\
\textbf{Contributing Authors' Emails:}\\ saidul.islam@concordia.ca;\\ jamal.bentahar@ku.ac.ae; \\rcohen@uwaterloo.ca;\\ grjoub@aut.edu.jo\\\\
\textbf{$^*$Corresponding Author's Email:}  jamal.bentahar@ku.ac.ae\\\\
The authors contributed equally to this work.
}

\usepackage{fancyhdr}
\renewcommand{\headrulewidth}{0pt}
\renewcommand{\footrulewidth}{0pt}
\fancypagestyle{first}{%
\fancyhf{} % clear all header and footer fields

\setlength{\headsep}{0.5in}
\renewcommand{\headrulewidth}{0pt}
\renewcommand{\footrulewidth}{0pt}}
\begin{document}

  \begin{center}
\maketitle
\thispagestyle{first}
\end{center}
\begin{abstract}

Cardiopulmonary resuscitation (CPR) is a critical, life-saving intervention aimed at restoring blood circulation and breathing in individuals experiencing cardiac arrest or respiratory failure. Accurate and real-time analysis of biomedical signals during CPR is essential for monitoring and decision-making, from the pre-hospital stage to the intensive care unit (ICU). However, CPR signals are often corrupted by noise and artifacts, making precise interpretation challenging. Traditional denoising methods, such as filters, struggle to adapt to the varying and complex noise patterns present in CPR signals. Given the high-stakes nature of CPR, where rapid and accurate responses can determine survival, there is a pressing need for more robust and adaptive denoising techniques. In this context, an unsupervised machine learning (ML) methodology is particularly valuable, as it removes the dependence on labeled data, which can be scarce or impractical in emergency scenarios. This paper introduces a novel unsupervised ML approach for denoising CPR signals using a multi-modality framework, which leverages multiple signal sources to enhance the denoising process. The proposed approach not only improves noise reduction and signal fidelity but also preserves critical inter-signal correlations (0.9993) which is crucial for downstream tasks. Furthermore, it outperforms existing methods in an unsupervised context in terms of signal-to-noise ratio (SNR) and peak signal-to-noise ratio (PSNR), making it highly effective for real-time applications. The integration of multi-modality further enhances the system's adaptability to various biomedical signals beyond CPR, improving both automated CPR systems and clinical decision-making.


\end{abstract}
\section{Introduction}\label{sec:intro}

\subsection{Problem Statement}

% CPR is a life-saving medical procedure that has been instrumental in the field of emergency medicine for several decades~\citep{hurt2005modern}. It started as a technique to prevent people from drowning but has now become a widely recognized procedure for anyone experiencing cardiac arrest~\citep{cooper2006cardiopulmonary}. The main goal of CPR is to keep the heart pumping and oxygenated blood continues to circulate to vital organs, especially the brain ~\citep{raza2021}. With more than 400,000 cases each year in North America and an average survival rate of $10\%$ \citep{meier2010chest}.  The American Heart Association and other global health organizations have emphasized the significance of CPR, not just among healthcare professionals but also among the general public~\citep{hinkelbein2020}. This is because the majority of cardiac arrests occur outside of hospital settings, where immediate medical intervention is not readily available. In these cases, CPR performed by bystanders can be crucial in saving lives~\citep{pellegrino2021}.

Cardiopulmonary resuscitation (CPR) is a vital, life-saving technique that can make a significant difference in medical emergencies by restoring blood flow and breathing in individuals experiencing cardiac arrest or respiratory failure. The primary objective of CPR is to keep the heart pumping and ensure continuous oxygenated blood flow to vital organs, particularly the brain ~\citep{raza2021}.  The American Heart Association, alongside other global health organizations, underscores the significance of CPR training, not just among healthcare professionals but also among the general public~\citep{hinkelbein2020}, as the majority of cardiac arrests occur outside of hospital settings, where immediate medical intervention may not readily available. In such scenarios, manual, human-driven CPR is predominantly practiced despite several difficulties and challenges that hinder successful CPR. For instance, performing CPR is physically demanding for individuals, even for health care providers for an extended period. Moreover, it is difficult to maintain the correct compression depth, rate, and consistency. Additionally, obtaining real-time feedback is significantly challenging, which is very crucial during the initial treatment in ambulances or hospital environments. To address these limitations, there has been a growing need for automated CPR systems over the last two decades, driven by advancements in mechanical devices within the healthcare system ~\citep{Contex00}. Automated CPR systems offer the potential to enhance the effectiveness of life-saving interventions by ensuring optimal compression and decompression at precise intervals for long periods. However, automating CPR faces several challenges related to decision-making and accuracy in compression and decompression. In this regard, Artificial intelligence(AI), particularly machine learning (ML), offers a promising way to address these limitations in medical operations. ML is capable of identifying complex patterns in biomedical data, improving the quality of CPR in medical systems, enhancing diagnostic accuracy, and enabling real-time monitoring of CPR in diverse medical conditions \citep{Contex02}. Furthermore, various biomedical signals associated with CPR—such as blood pressure, velocity, force, and respiratory waveforms—play a critical role in guiding treatment decisions, particularly in intensive care units (ICUs). Accurate monitoring and processing of these signals during CPR is essential for assessing patient response, informing medical interventions, and ultimately improving patient outcomes \citep{hurt2005modern}.\\

CPR signals are inherently dynamic and time-sensitive, requiring specialized approaches for effective processing and decision-making. Moreover, these signals are often corrupted by various types of noise sources, including baseline wander, electrode artifacts, motion artifacts, and electromagnetic interference, which significantly degrade signal quality and complicate interpretation \citep{limitation02}. Ensuring the integrity and quality of signal data is crucial for effective ML approaches in accurate decision-making for CPR automation and monitoring systems. While traditional signal processing techniques such as filtering, offer some improvements, they often struggle to remove noise effectively without compromising essential physiological features \citep{limitation00}. Since CPR is a process of emergency healthcare, where decision-making can directly impact survival, it is unacceptable to compromise the quality of the signal. Therefore, achieving effective denoising while preserving key signal characteristics is essential. Moreover, one limitation of conventional filters lies in their dependence on predefined parameters—such as cutoff frequencies or filter orders—and assumptions regarding noise characteristics, which may not correspond to the complex and dynamic nature of real-world CPR signals. Traditional filters often struggle to effectively remove noise without inadvertently distorting or attenuating important signal components, leading to the potential loss of valuable information. Additionally, filters are generally tailored to specific noise types or signal modalities, limiting their adaptability to diverse noise profiles and environments \citep{limitation01}. These inherent constraints underscore the need for more flexible and adaptive denoising techniques capable of preserving critical signal features. Given the high-stakes nature of CPR, where accurate signal interpretation is a matter of life and death, overcoming these limitations is essential for advancing automated CPR systems and improving patient outcomes. \\

The application of machine learning (ML) has emerged as a promising approach to tackling critical tasks in medical systems. ML excels at identifying complex patterns and relationships within biomedical data, offering solutions that can enhance diagnostic accuracy, improve healthcare systems, and enable real-time monitoring across a variety of medical conditions \citep{challange05}. Within the domain of CPR, integrating ML presents unique opportunities, as CPR signals are both dynamic and time-sensitive, requiring specialized approaches for effective processing and decision-making. ML-based approaches, particularly for tasks such as denoising and signal enhancement, demonstrate the ability to learn from diverse, heterogeneous, and noisy data, thus showing great potential in deciphering the intricate nuances inherent in CPR signals. Moreover, recent advancements in computational power, particularly with the advent of high-performance GPUs and cloud computing platforms, have significantly improved the feasibility of deploying ML algorithms in real-time applications, thus facilitating timely insights in critical medical scenarios such as CPR \citep{MLcomputation}.  Such advancements can substantially improve the accuracy and reliability of real-time monitoring during life-saving interventions. However, despite these promising prospects, applying ML algorithms to CPR signal denoising presents several notable challenges. These include the development of robust feature extraction methodologies that are tailored to the unique characteristics of CPR signals, selecting appropriate ML frameworks and models capable of processing temporal data and noise, and acquisition of labeled datasets for supervised learning tasks \citep{limitation03}. Existing ML approaches for CPR signal denoising are predominantly supervised, relying on clean labeled data that corresponds to noisy signals. In real-world scenarios, obtaining such labeled data is extremely difficult, if not impractical, thereby limiting the applicability of supervised ML methods. Consequently, there is a clear need for an unsupervised ML approach dedicated to CPR signals that can effectively denoise CPR signals without reliance on labeled data. Currently, no unsupervised ML methodology specifically designed for CPR signal denoising has been developed, leaving a gap in the existing research landscape. \\

% Despite these challenges, very few ML approaches can be observed that are supervised which means, the approaches need corresponding clean labeled data. In a real-life context, labeled clear data corresponding to noisy data is unavailable or very difficult to achieve, so the implementation of a supervised ML approach for CPR denoising is very difficult. In this context, a dedicated unsupervised ML method is necessary that can denoise CPR signal without any help of labeled data. Whereas, a dedicated unsupervised ML method for CPR signal denoising is still missing in the research domain.  \\

\subsection{Novelty and Contributions}

In this paper, we address the challenges by introducing an advanced method for CPR signal denoising and artifact removal, aimed at enhancing patient outcomes and healthcare delivery. A novel unsupervised ML methodology has been proposed, specifically designed for denoising biomedical signals during CPR. The framework introduces a multi-modality approach, enabling the concurrent processing of multiple signals while addressing the noise characteristics of each signal individually. This individualized signal processing facilitates the extraction of unique features, thereby improving the overall denoising process and enhancing the accuracy of signal interpretation. A key feature of the proposed framework is its use of unsupervised ML, which demonstrates the method's capability to effectively remove noise and enhance signal fidelity without the need for labeled data. By identifying underlying structures and patterns in noisy CPR signal data, this approach is particularly well-suited for real-world applications, where labeled ground truth is often scarce or unavailable. The key contributions of this paper are as follows: (i) a comprehensive exploration of existing denoising approaches for CPR signals, (ii) the introduction of a novel multi-modal ML methodology that processes multiple signals simultaneously, (iii) the use of an unsupervised ML approach, eliminating the dependency on labeled data, (iv) the generation of synthetic biomedical CPR signals for research purposes, and (v) the conceptualization of a multi-modal ML framework for enhanced decision-making in medical emergencies.\\

Moreover, a comprehensive comparison of the proposed method has been conducted with existing ML and filter methods in an unsupervised context to justify the performance of the proposed methodology. The effectiveness of the methodology is demonstrated through visualizations, as well as signal-to-noise ratio(SNR) and peak signal-to-noise ratio(PSNR) scores. Notably, the method successfully preserves the signal data correlations at a significant level, even while processing each signal through dedicated models and subsequently combining them. To illustrate this, the correlations matrix before and after the denoising process have been calculated and quantified the differences by determining the correlation coefficients, which are crucial for downstream tasks. The results highlight the potential of this approach to enhance biomedical signal processing not only in CPR scenarios but also in border medical contexts. The paper is organized into six sections, the current section discusses the context and motivation behind this paper, while Section \ref{Sec:related_task} presents a review of related literature, highlighting the limitations of existing methods and differentiating our contributions. Section \ref{sec:Preliminaries} outlines the preliminary concepts essential for the proposed methodology. The details of the methodology and proposed solution are presented and depicted in Section \ref{sec:Methodology}. Importantly, Section \ref{Sec:result} showcases the output results compares the findings with the existing methods, and extends the discussion to data correlation preservation. Section \ref{future_work} presents a discussion addressing the scope of explainability, and exploring the application of the multi-modality concept in decision-making processes that integrate multiple sources of information during medical emergencies. Finally, Section \ref{conclusion} summarizes the key findings and contributions of this paper.
%
% \begin{itemize}
% \item Proposed a multi-modality ML approach.
% \item Applied an unsupervised methodology.
% \item Generate biomedical CPR signals from the Babbas model.
% \item Provides enhanced signal quality and preserves signal correlation.
% \end{itemize}

\section{Related Work}\label{Sec:related_task}

This section explores the landscape of denoising techniques for biomedical signals, specifically focusing on CPR signals. While several ML-based methods for denoising electrocardiogram (ECG) signals have been widely researched, the consideration of ECG signal denoising techniques has been intentionally set aside due to the significant differences between ECG and CPR signals. Unlike ECG signals, which are electrical recordings of the heart's activity, CPR signals are mechanical signals generated during chest compressions. Additionally, ECG signals exhibit characteristic waveforms representing different phases of the heart's electrical activity, while CPR signals exhibit even more complex patterns and instead represent the rhythmic application of pressure to the chest. Furthermore, ECG signals are typically measured using electrodes placed on the skin surface and analyzed for cardiac rhythm abnormalities \citep{method_related02}. On the contrary, CPR signals are measured using external sensors during CPR maneuvers to monitor chest compression effectiveness. Importantly, CPR signals involve multiple signals from different body parts, including the abdomen, necessitating dedicated fashion denoising methods due to their unique characteristics \citep{method_pre}. Consequently, there is an urgent need for robust and automated denoising techniques using a dedicated methodology that can enhance the quality of CPR signals by removing artifacts and facilitating accurate clinical decision-making. \\

\renewcommand{\arraystretch}{2}
\begin{longtable}{|M{2.5cm}|M{2cm}|M{5cm}|M{5cm}|}
    \hline
    \textbf{Approach} & \textbf{Type of Method} & \textbf{Limitations} & \textbf{Differences} \\
    \hline
    \endfirsthead
    
    \multicolumn{4}{c}{{\bfseries \tablename\ \thetable{} -- Continued from previous page}} \\
    \hline
    \textbf{Approach} & \textbf{Type of Method} & \textbf{Limitations} & \textbf{Differences} \\
    \hline
    \endhead
    
    \hline \multicolumn{4}{|r|}{{Continued on next page}} \\ \hline
    \endfoot
    
    \endlastfoot

    \cite{related_05} & Filter & The denoising method is proposed for a specific frequency of the signal &  Our methodology can handle different range of frequency of biomedical signals with the flexibility and efficiency of signal denoising, \\
    \hline
    
    \cite{related_02} & Filter (Pursuit-like approach) & Performance depended and highly sensitive to filter design parameters like a fixed sample rate, window samples, and number of channels which might not be effective for all signals. & Our machine learning-based framework is more flexible for signal variations while modifications can bring better performance.   \\
    \hline
    
   % \cite{related_03} & Machine learning & Capable of processing only one type of data i.e ECG signal from the CPR & Our methodology can process multiple numbers of cpr associated signals at the same time, while different types of model can be used based on signal type and needs. \\
   %  \hline
    
    \cite{related_04} & Filter & Performance is sensitive and depends on the characteristics of the currpted signal. &  Our machine learning-based framework is more flexible for signal variations while modifications can bring better performance.\\
    \hline
    
    \cite{related_06} & Filter & The performance varies a lot based on the signals of various conditions during CPR & Our proposed methodology is capable of handling a variety of signals, while the method can tackle different signals according to their merits. \\
    \hline

    \cite{related_01} & Filter (Enhanced adaptive filter) & Uncertain about the performance during resuscitation due to lack of capability to identify variation in CPR artifacts induced by manual versus mechanical chest compressions.  & Our method can handle diverse types of biomedical signals with varying noise characteristics and enhances the flexibility and efficiency of signal denoising, \\
    \hline
    
    \cite{related_07} & Machine learning & Need a corresponding clean signal for this method and applicable only for specific CPR signals, not diverse to all types of CPR signals. & Our proposed methodology can handle various signals in an unsupervised manner (without corresponding clean signal), while the method can tackle different signals according to their merits. \\
    \hline
    
    \caption{\label{tab: related_works}Comparative summary between our proposed ML methodology and the existing CPR signal denoising methods.}
\end{longtable}

This part of the paper thoroughly explores existing denoising techniques designed specifically for biomedical signals, with a particular focus on CPR signals. The goal is to identify research papers proposing novel models and methods tailored for effectively denoising CPR signals. However, few studies have focused on CPR signal denoising, in contrast to the extensive research available on ECG signal denoising. Despite this limitation, the review of the existing literature uncovered a diverse array of denoising techniques, including traditional signal processing methods such as filtering. The applicability of these methods to CPR signals has been investigated, revealing significant limitations in existing approaches. Table \ref{tab: related_works} summarizes the differences and advantages offered by our methodology examined and compared to these existing techniques. For instance, \cite{related_05} introduced a filter-based method targeted at removing artifacts from CPR signals by operating within a specific frequency range. Similarly, \cite{related_02} experimented with a pursuit-like filtering approach for CPR signal denoising with limited applicability across diverse CPR signal types and specific filtering parameters. The work by \cite{related_04} and \cite{related_06} also introduced filter-based techniques, but their performance was found to be heavily dependent on signal characteristics and conditions. Additionally, \cite{related_01} developed an enhanced adaptive filter for CPR signal denoising, but its efficacy remained uncertain due to challenges in capturing signal variations adequately. Lastly, \cite{related_07} introduced a machine learning-based supervised approach handling CPR signal artifacts, however, it addressed the limitation in applicability which is only for specific CPR signals and not being applicable across diverse CPR signal types remains a challenge, given the inherent variability in signal patterns and characteristics. These findings underscore the pressing need for more robust and adaptable denoising methodologies that can effectively handle the variability of all CPR signal types. Such advancements are crucial for improving signal fidelity and enabling accurate clinical decision-making in real-time settings.

%\medskip

\section{Preliminaries}\label{sec:Preliminaries}
% In developing our method for denoising CPR signals, we recognized the unique characteristics of these biomedical signals and the need for specialized processing techniques to address their complexities effectively. Our approach is founded on the integration of advanced machine learning (ML) components that are specifically chosen to suit the dynamic and time-sensitive nature of CPR signals. We considered factors such as the presence of various forms of noise and interference, aiming to optimize signal processing capabilities. To achieve this, we combined different ML techniques, including autoencoders, CNNs, residual connections, and most importantly multi-modality concepts. These components form the backbone of our approach, enabling us to extract important signal features, remove unwanted noise, and enhance signal clarity. In this section, we will provide a comprehensive explanation of each component in simple terms and demonstrate their potential contributions to the development of our proposed framework for CPR signal processing.\\
The proposed approach uses advanced ML concepts and techniques that are carefully selected to address the dynamic and urgent nature of CPR signals. The framework considers various types of noise and interference that can compromise the integrity of these signals, such as baseline wander, motion artifacts, and electromagnetic interference, and aims to enhance the overall signal quality during processing. The core of this method integrates a multi-modality concept alongside several ML techniques, including autoencoders, convolutional neural networks (CNNs), and residual connections. Each component plays a crucial role in forming an unsupervised CPR signal processing framework.\\

\subsection {Multi-modality approach}

% In general, Multi-modal deep learning aims to leverage information from diverse modalities, such as text, image, and audio, to enhance model performance. Initially, each modality-specific data undergoes separate processing using specialized neural network architectures tailored to their characteristics. These models extract high-level features and encode them into lower-dimensional representations. Once features from each modality are extracted, they are fused to create a joint representation, achieved through early fusion, where raw input data from different modalities are combined before processing, or late fusion, where features are combined at a later stage using various techniques. The fused representation is then fed into a joint learning model, which integrates information from all modalities to make predictions or perform tasks. This joint model can range from simple concatenation followed by fully connected layers to more complex architectures. Training the entire multi-modal architecture involves end-to-end optimization using back propagation and gradient descent. Parameters of modality-specific models and fusion layers are updated simultaneously to minimize the loss function, ensuring alignment between predicted and true labels or outputs. By effectively integrating information from multiple modalities into a joint representation, multi-modal deep learning models capture richer and more comprehensive information, leading to enhanced performance in executing a range of tasks.\\

\begin{figure}[ht]
    \begin{center}
    \includegraphics[width=.65\linewidth,height=.27\textwidth]{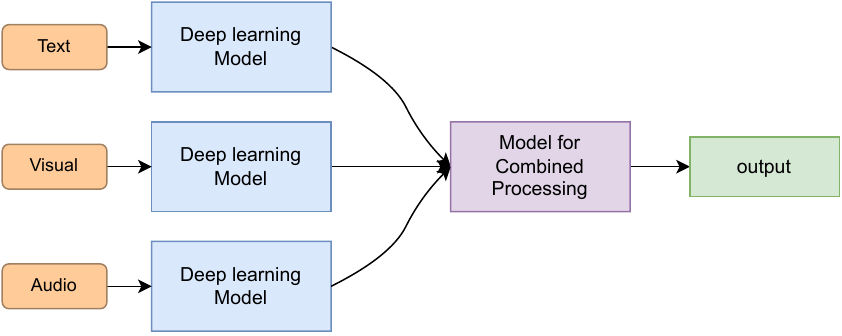}
    \caption{ General Multi-Modality approach for different types of data.}
    \label{fig:multi-model}
    \end{center}
    %\vspace{-12pt} % reduce the space after the picture by 10pt
\end{figure}

There are several motivating factors for choosing a multi-modality approach for effective CPR signal processing. First, it holds substantial promise for improving healthcare effectiveness by integrating various types of medical data, leading to more comprehensive, accurate, and personalized patient assessments \cite{method_multi_000}. In addition, multi-modality can enhance model transparency and the overall explainability of AI applications in healthcare, opening doors to a broader range of clinical applications and making outcomes more interpretable for practitioners \cite{method_multi_00, witold_}. In general, multi-modal deep learning is an approach that integrates information from various types of data, like text, images, and audio, to enhance model performance and accuracy. In this method, each data modality is initially processed independently using specialized neural networks tailored to the specific characteristics of that modality, as illustrated in Figure \ref{fig:multi-model}. These networks extract high-level features from the data and encode them into lower-dimensional representations, capturing essential patterns while reducing noise and redundancy. Once features from each modality are extracted, they are fused into a single, unified representation. This can happen early on when the raw data is combined before processing to learn cross-modal relationships, or later when the processed features are combined using different methods \citep{method_multi0}.\\

Once the combined representation is formed, a unified model utilizes it to make predictions or perform downstream tasks. The architecture of this joint model can vary in complexity, ranging from simple concatenation followed by fully connected layers to more complex architectures. During training, the model's parameters—including those of the modality-specific networks and the fusion layers—are updated simultaneously through backpropagation. This ensures that all components work in harmony, optimizing the overall system to minimize the loss function and align predictions with the true labels or outputs. By integrating information from different modalities, multi-modal deep learning allows the model to capture richer, more nuanced representations of the data. This, in turn, enables the system to handle more complex tasks and improves its ability to generalize, making it particularly well-suited for applications where diverse data sources are available, such as medical diagnosis or real-time monitoring in critical settings \citep{method_multi03}.

% In our framework, we capitalize on the multi-modality concept to process multiple signals simultaneously while denoising each signal separately before amalgamating them into a unified input data shape. This approach confers several advantages. Firstly, by processing multiple signals concurrently, we can harness correlations and dependencies between different signals to enhance denoising performance. Each signal contributes complementary information, leading to improved signal quality and fidelity. Secondly, denoising each signal individually enables a tailored approach to address the unique characteristics and noise profiles of each signal, resulting in more effective noise removal and signal enhancement. Additionally, this method offers flexibility and adaptability to handle diverse signal modalities and noise sources, making it applicable to a wide range of biomedical signal denoising tasks. Moreover, the scalability of our approach allows for efficient handling of complex denoising tasks, even in scenarios with large volumes of data or high-dimensional input. Finally, by denoising each signal separately before combining them, our framework mitigates the impact of noise variations across different signals, ensuring robustness to noise fluctuations and consistent denoising performance. Overall, our approach offers a comprehensive and effective solution for biomedical signal denoising, facilitating more accurate clinical interpretation and decision-making.

\subsection{Autoencoder}

% An autoencoder, a type of artificial neural network, serves the purpose of efficiently representing data by minimizing the difference between its output (reconstructed data) and the clean input data. Notably, autoencoders can be applied in both supervised and unsupervised manners. In the supervised technique, pairs of noisy input data and corresponding clean data are fed into the autoencoder for training. Conversely, the unsupervised approach involves training the autoencoder solely with noisy input data, without corresponding clean data. Throughout the training process, the autoencoder learns to encode noisy input data into a latent representation, capturing the underlying structure and patterns of the data.\\

An autoencoder is a type of artificial neural network designed to efficiently represent data by making its output (reconstructed data) as close as possible to the clean input data. It can be used in two main ways: supervised and unsupervised. In the supervised method, pairs of noisy input data and their corresponding clean versions are given to the autoencoder for training. However, in the unsupervised approach, only noisy input data is used for training, without any clean data. During training, the autoencoder learns to encode the noisy input data into a hidden representation, capturing the important patterns and structure of the data \citep{method_anco00}.\\

\begin{figure}[htbp]
    \begin{center}
    \includegraphics[angle=90,width=.6\linewidth,height=.27\textwidth]{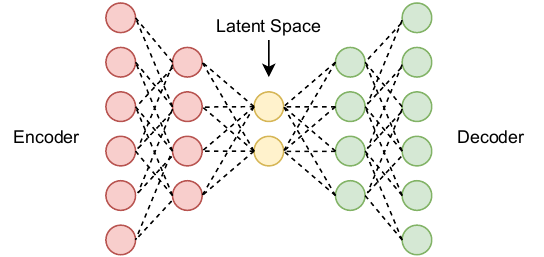}
    \caption{ Vanilla Autoencoder model.}
    \label{fig:Autoencoder_Diagram}
    \end{center}
    %\vspace{-12pt} % reduce the space after the picture by 10pt
\end{figure}

% The autoencoder comprises two essential components: the encoder and the decoder. The encoder plays a pivotal role in transforming input data samples into a latent representation. Utilizing neural network architectures such as recurrent neural networks (RNNs), Long Short-Term Memory (LSTM) networks, CNNs, and so on, the encoder extracts high-level features from the input data. Its primary objective is to learn a condensed representation of the input that encapsulates its crucial characteristics while effectively filtering out noise. By capturing both spatial and temporal dependencies within the data, the encoder ensures that essential information is retained while noise is appropriately suppressed. On the other hand, the decoder component of an autoencoder is responsible for reconstructing output data from the representations generated by the encoder. Collaborating closely with the encoder, the decoder works to reverse the dimensionality reduction process implemented during encoding, thereby restoring the original structure and features of the input data, and translating them back into high-dimensional output data. Through iterative reconstruction, the decoder aims to minimize the reconstruction error between the output and the original input data, ensuring fidelity to the underlying signal while effectively suppressing noise artifacts.

The autoencoder consists of two main components: the encoder and decoder. The encoder takes input data and transforms it into a compact representation called a latent representation, as shown in Figure \ref{fig:Autoencoder_Diagram}. It uses neural network architectures like recurrent neural networks (RNNs), LSTM networks, CNNs, etc., to extract important features from the input data. The goal of the encoder is to create a condensed version of the input that captures its important characteristics while removing noise. It considers both spatial (how data points relate to each other in space) and temporal (how data changes over time) aspects of the input to ensure that vital information is preserved while noise is filtered out effectively. On the other hand, the decoder part of the autoencoder works alongside the encoder. Its job is to reconstruct output data from the latent representations produced by the encoder. By collaborating closely with the encoder, the decoder reverses the process of dimensionality reduction that occurred during encoding. This means it restores the original structure and features of the input data, translating them back into high-dimensional output data. Through a series of iterations, the decoder aims to minimize the difference between the output and the original input data, ensuring that the reconstructed data closely resembles the original signal while effectively removing noise \citep{method_anco02}.

\subsection{Convolution neural network}

One-dimensional Convolutional Neural Networks (1D CNNs) are essential components in deep learning architectures, known for their ability to capture intricate complex patterns and spatial relationships within sequential data. A typical 1D CNN consists of three primary components: convolutional layers, pooling layers, and fully connected layers, each serving a distinct function. In the convolutional layer, learnable filters are applied to input data using convolution operations, generating feature maps that encode local patterns \citep{method_cnn04}. These filters learn to detect specific features by scanning across the input data's spatial dimensions. Pooling layers are responsible for reducing the size of feature maps while preserving important information. Techniques like max pooling or average pooling aggregate feature map values within localized regions, help to summarize feature presence across different areas and improve computational efficiency. Finally, fully connected layers combine extracted features from previous layers, allowing the network to perform specific tasks. These layers establish connections between all neurons in adjacent layers, facilitating the representation of high-level features and making predictions tailored to the task at hand \citep{method_cnn02}. Mathematically, the 1D CNN can be represented as follows: Given an input sequence $X_{\text{input}}$ with dimensions $(N, L, 1)$, where $N$ is the number of samples, $L$ is the sequence length (number of time steps), and $1$ represents the single feature dimension, the output $X_{cnn}$ of the 1D CNN component can be obtained as follows:

\begin{equation}
X_{cnn} = Conv1D(X_{input}, filters=n, kernel\_size=n, activation=ReLU)
\end{equation}

\noindent where $Conv1D$ denotes the 1D convolution operation with $n$ number of filters and a kernel size of $n$. activation function is applied to introduce non-linearity, here for instance $ReLU$ (Rectified Linear Unit).\\\\
Next, a $MaxPooling1D$ layer with a pool size of $n$ is applied to downsample the data and reduce the computational complexity:

\begin{equation}
X_{cnn} = MaxPooling1D(X_{cnn}, pool\_size=n)
\end{equation}

\vspace{0.45cm}

To capture more intricate patterns, more 1D CNN layers can be added, comprising $n$ filters with a kernel size of $n$ and activation function. Furthermore, more $MaxPooling1D$ layers also can be applied to further downsampling the data \citep{method_cnn03}.

\section{Methodology and Proposed Solution\label{sec:Methodology}}
\subsection{Overview and Framework}

\begin{figure}[ht]
    \begin{center}
    \includegraphics[width=1\linewidth,height=.85\textwidth]{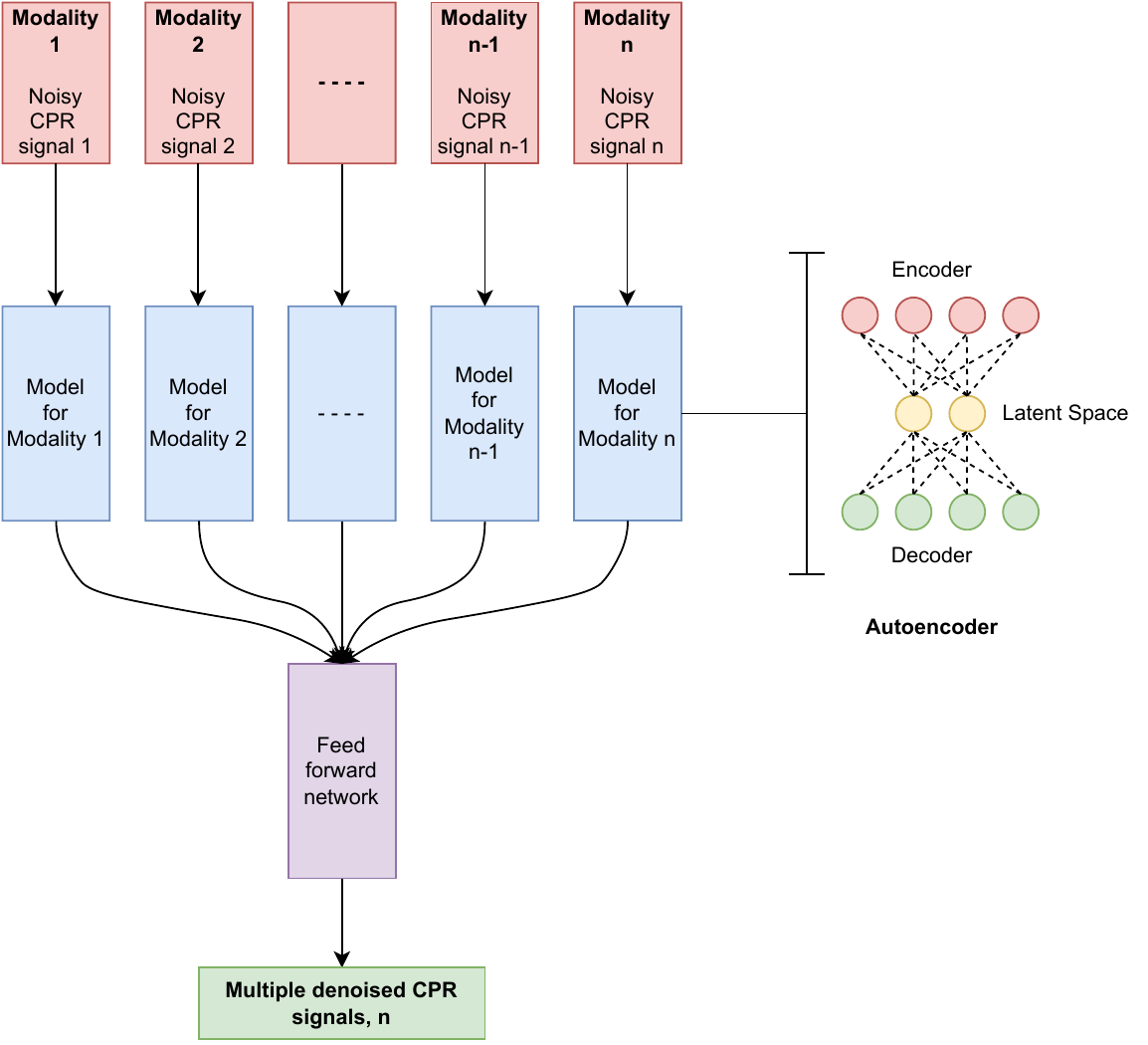}
    \caption{ Proposed Multi-modal Machine Learning Framework for Denoising Biomedical Signal during CPR.}
    \label{fig:framework}
    \end{center}
    \vspace{-12pt} % reduce the space after the picture by 10pt
\end{figure}

CPR signals are highly correlated, presenting challenges in automating their processing. Traditional ML methods treat signals collectively, but denoising while preserving their interrelationships is difficult due to each signal's unique characteristics \citep{method_fram00}. To address this, we propose an ML-based methodology that denoises each signal individually while maintaining their correlations, improving accuracy and reliability for downstream tasks like real-time monitoring. In the proposed framework, the concept of multi-modality is leveraged to simultaneously process multiple signals while individually denoising each signal before amalgamating them into a unified input data shape, as depicted in Figure \ref{fig:framework}. Each signal is treated as a distinct modality, and accordingly, processed with a separate model tailored to its unique characteristics and requirements. This approach ensures that each signal is effectively denoised while preserving its individual attributes. After the separate processing of each signal with dedicated models, the outputs are fed through a feed-forward network, where they are combined into a unified input. The feed-forward network facilitates the integration of the denoised signals into a coherent and cohesive format, resembling the original noisy input data but in a denoised form. Subsequently, the denoised output from the feedforward network is prepared for downstream tasks or feature utilization in machine learning applications aimed at CPR automation.\\

The selection of an appropriate model for each modality or signal is a critical aspect in achieving effective denoising of CPR signals. Given the primary objective of this framework—denoising—the autoencoder model is employed due to its flexibility for use in both supervised and unsupervised learning settings. While the supervised approach involves training the autoencoder with pairs of noisy input data and corresponding clean data, the unsupervised method is chosen for considering our real-life scenario. In this unsupervised setting, the autoencoder learns to encode noisy input data into a latent representation, capturing underlying structures and patterns without relying on labeled clean data and then generating new data from the latent space using the decoder. Several architectural variations of autoencoders exist, such as RNNs, LSTMs, CNNs, and attention-based models \citep{method_fram}. Given the need to capture local features and complex patterns in the CPR signals, CNN-based denoising autoencoders have been adopted across all modalities. CNNs are particularly well-suited for this task due to their capacity for effective feature extraction and their robustness in handling noise variations, making them ideal for processing biomedical signals. The standardized use of CNN-based autoencoders ensures consistent performance across different signals. By leveraging CNNs' strengths in capturing spatial dependencies and hierarchical patterns, the approach is capable of demonstrating promising efficacy in mitigating noise artifacts and enhancing the fidelity of biomedical signals for improved analysis and interpretation \citep{method_fram_}.\\

% While different signals may benefit from specific architectural features, applying the same architecture-based autoencoder for each modality or signal is not essential in extensive application. However, it’s essential to recognize that certain signals may inherently benefit from alternative architectures-based autoencoders, which excel in capturing temporal dependencies or focusing on specific regions of interest for each signal.  Although the framework employs a uniform CNN-based approach, future iterations could explore hybrid architectures or adaptive model selection strategies to accommodate diverse signal characteristics beyond CPR signals better

\begin{figure}[ht]
    \begin{center}
    \includegraphics[angle=90, width=1\linewidth,height=.58\textwidth]{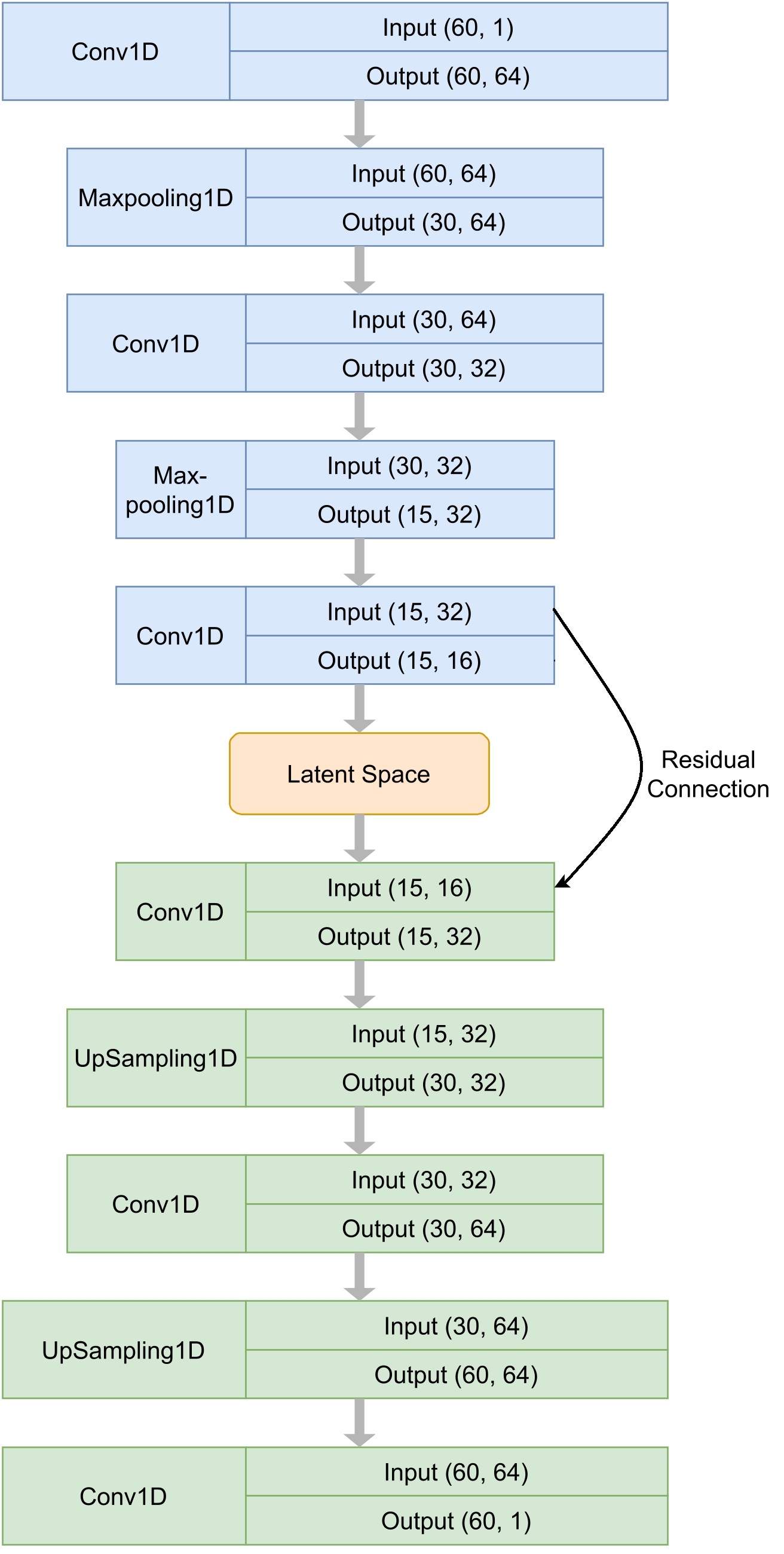}
    \caption{ Used Residual connected CNN-based denoising autoencoder in Multi-modal Machine Learning Framework.}
    \label{fig:CNN_autoencoder}
    \end{center}
    % \vspace{-12pt} % reduce the space after the picture by 10pt
\end{figure}

The CNN-based autoencoder model in the proposed framework enables the network to automatically discern relevant features across various spatial scales, enhancing its ability to capture intricate patterns within the data. In our autoencoder, two(2)layers of 1-dimensional CNN are deployed in both the encoder and decoder. After each of the 1D-Convolution layers, a max-pooling-1D layer has been placed in the encoder to compress the signals into latent space, as in Figure \ref{fig:CNN_autoencoder}. However, upsampling 1D layers have been used in the case of the decoder to generate signals from the latent space. One notable enhancement in our autoencoder is the incorporation of a residual connection. This connection involves introducing an additional convolutional layer after the last max-pooling-1D layer in the encoder, which captures supplementary features or representations from the encoder's output \citep{method_res02}. Subsequently, the output of this residual layer is combined with the output of the last convolutional layer of the encoder. This combined output is then passed through upsampling layers to prepare it for concatenation with the input of the decoder. By incorporating the residual connection in this manner, the autoencoder is enabled to leverage the information captured by the residual layer, enhancing its denoising capabilities. This process facilitates the propagation of relevant features while minimizing the impact of noise, thereby enabling the autoencoder to learn more robust representations and achieve superior denoising performance, particularly in deeper architectures. Ultimately, the inclusion of residual connections contributes to a clearer and more accurate reconstruction of the input signal, leading to improved signal quality \citep{method_res}.\\

\subsection{Methodology Implementation} \label{Sec:application}

Accessing real-world medical data, particularly CPR data poses significant challenges due to privacy concerns and regulatory constraints. These limitations make it difficult to utilize actual patient data for machine learning (ML) research in the CPR context. Despite an extensive search for publicly available CPR datasets, all identified datasets are private and restricted, creating a significant barrier to progress in this field. As a result, alternative data sources are necessary to advance research and development. To overcome this obstacle, the generation of synthesized data has emerged as a viable solution. Simulated data allows for experimentation, algorithm development, and validation while avoiding privacy issues and regulatory hurdles. The Babbs model, a well-established mathematical model for simulating CPR processes, was identified as the most suitable framework for this task \citep{babbs,babbs__}. This model is renowned for its ability to accurately replicate the complex dynamics of CPR, including variations in signal morphology and the introduction of realistic noise. By leveraging the Babbs model, realistic CPR scenarios are simulated, enabling robust research into CPR signal processing, monitoring systems, and decision-making algorithms without compromising patient confidentiality \citep{babbs00}.

\subsubsection{Data Generation} 
% The Babbs model is a comprehensive mathematical model that describes the physiological changes during CPR. It incorporates various factors, including chest compression depth (D), compression rate (R), and duty cycle (Duty), as well as their effects on coronary perfusion pressure (CPP) and other physiological parameters. The Babbs model is based on extensive research and has been validated in experimental and clinical studies. \cite{}

The Babbs model is a comprehensive mathematical framework that describes the physiological changes occurring during CPR. It allows for real-time calculations of coronary perfusion pressure (CPP) based on chest compression parameters. The model can mimic the dynamic changes in the elastance and resistance of the arterial system during CPR, influenced by factors such as chest compression depth, rate, and duty cycle \citep{babbs01}. This provides a quantitative method to optimize these parameters to achieve the desired CPP, which provides improved perfusion during CPR. Specifically, the Babbs model incorporates variables like chest compression depth (D), compression rate (R), and duty cycle (Duty), and analyzes their effects on CPP and other physiological parameters. It has been proven in extensive research and validated through experimental and clinical studies. All of these factors make the Babbs model a reliable tool for simulating CPR scenarios and optimizing resuscitation techniques \citep{babbs02}.\\

The Babbs model equations can be summarized as follows:

\begin{align}
CPP &= \frac{DBP \times Duty}{Duty + \frac{1}{R}} \\
DBP &= E \times D - F \times D^2 \\
E &= E\min + \frac{E\max - E\min}{1 + exp(PE \times (Dtarget - D))} \\
F &= F\min + \frac{F\max - F\min}{1 + \exp(PF \times (Dtarget - D))}
\end{align}

\vspace{0.45cm}

where the variables are as follows: $CPP$: Coronary perfusion pressure during CPR;
$DBP$: Diastolic blood pressure during CPR;
$D$: Chest compression depth;
$R$ : Chest compression rate;
$Duty$: Chest compression duty cycle;
$E$: Elastance of the arterial system;
$F$: Resistance of the arterial system;
$E\min$, $E\max$: Minimum and maximum elastance values;
$F\min$, $F\max$: Minimum and maximum resistance values;
$PE$, $PF$: Model parameters;
$Dtarget$: Target chest compression depth.\\

% The Babbs model allows for real-time calculations of CPP based on the chest compression parameters. It takes into account the dynamic changes in the elastance and resistance of the arterial system during CPR, which are influenced by the chest compression depth, rate, and duty cycle. This model provides a quantitative way to optimize chest compression parameters to achieve the desired CPP for improved perfusion during CPR.

\begin{figure}[htbp]
    \begin{center}
    \centering
    \includegraphics[width=1\textwidth,height=.28\textwidth]{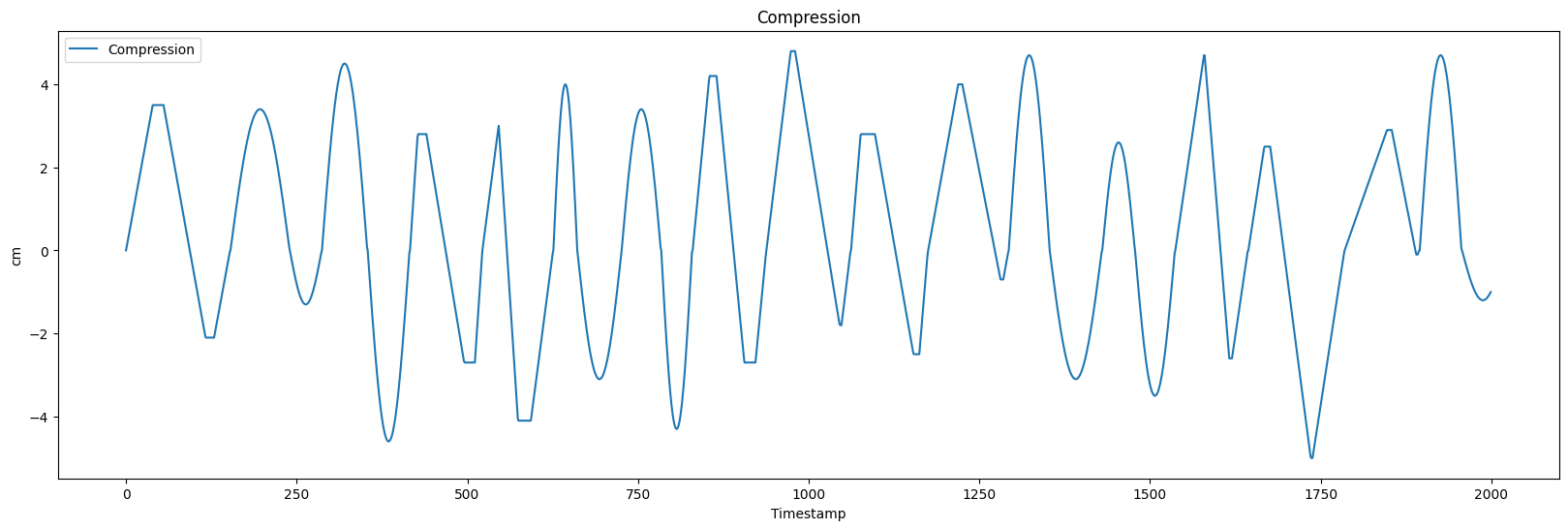}
    \includegraphics[width=1\textwidth,height=.28\textwidth]{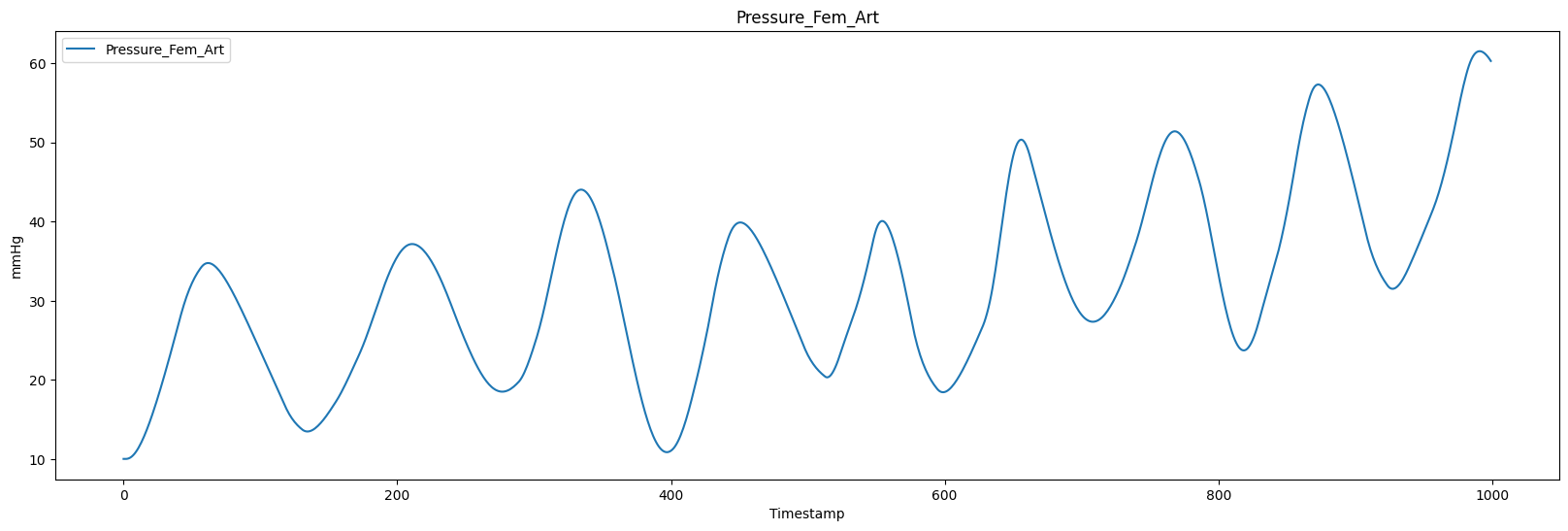}
    \includegraphics[width=1\textwidth,height=.28\textwidth]{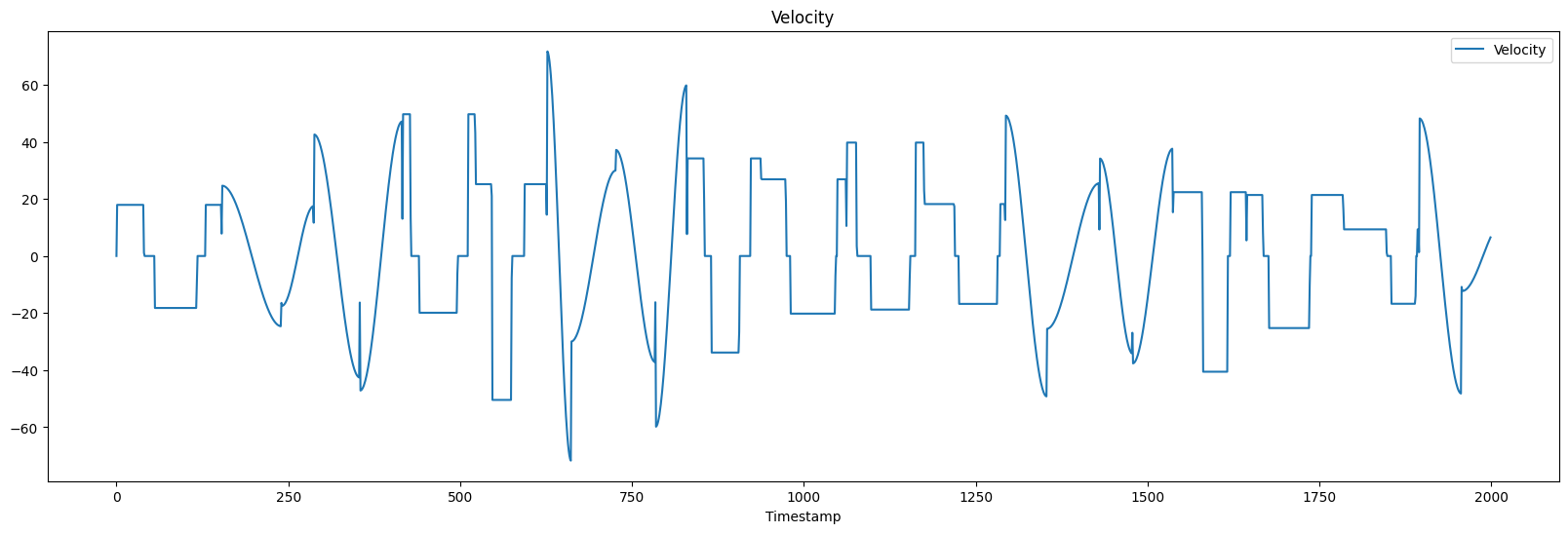}
    \includegraphics[width=1\textwidth,height=.28\textwidth]{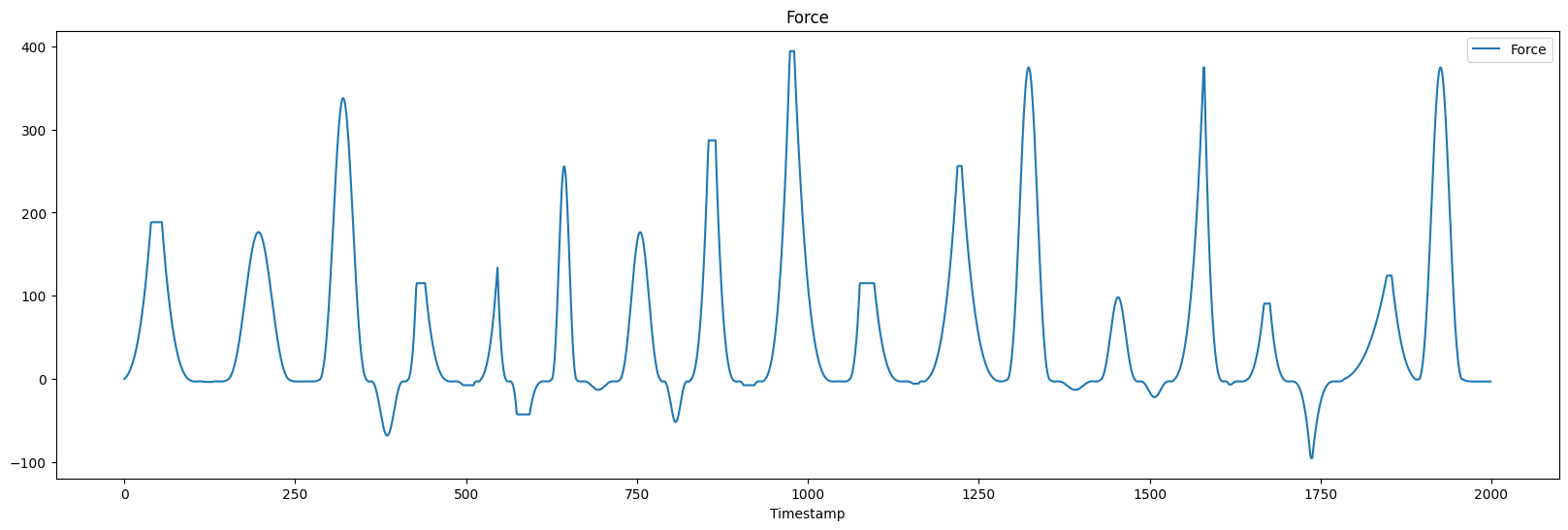}
    \includegraphics[width=1\textwidth,height=.28\textwidth]{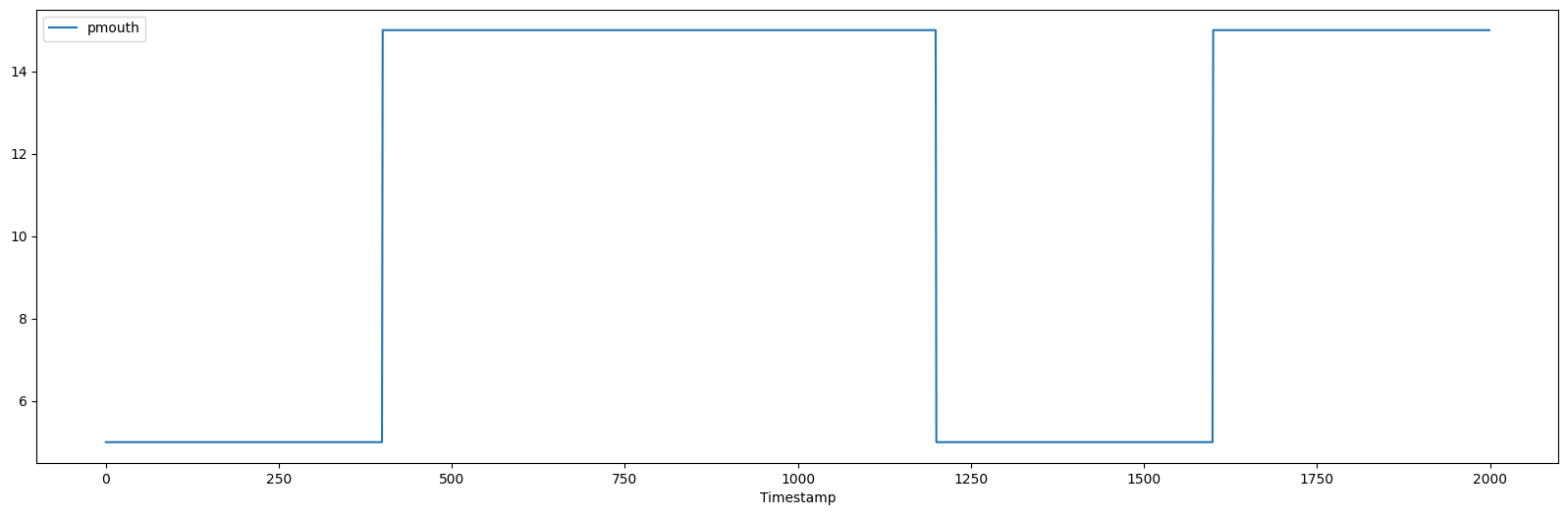}
    \caption{Generated CPR- Compression, Pressure, velocity, Force, Pmouth signals.}
    \label{fig: clean_data.}
    \vspace{-5pt} % reduce the space after the picture by 10pt
    \end{center}
\end{figure}

The Babbs model optimizes chest compression parameters during CPR by providing real-time feedback to guide rescuers in achieving optimal chest compressions. The algorithm implementation involves measuring compression depth, duty cycle, and rate during CPR and using these values to calculate CPP based on the Babbs model equations. The calculated CPP is then compared to the optimal CPP target range, and the algorithm provides feedback to the rescuers to adjust compression parameters accordingly. A total of $100$ cycles of chest compressions is simulated with a compression rate of $100$ compressions per minute. The compression depth was set to $50$ mm, and the decompression depth was set to $10$ mm. The time of compression and time of decompression were set to $50\%$ and $50\%$ of the total cycle time, respectively. The target diastolic blood pressure was set to $40$ mmHg. To generate the synthetic data for various patients, the external force is varied and applied during chest compressions, chest compliance, and airway resistance to simulate different patient conditions. The external force was varied from $500$ N to $1000$ N in steps of $100$ N, however, Chest compliance was varied from $0.01$ L/kPa to $0.05$ L/kPa in steps of $0.01$ L/kPa. Moreover, Airway resistance was varied from $1$ cmH2O/(L/s) to $5$ cmH2O/(L/s) in steps of $1$ cmH2O/(L/s). The potential clinical implications of this approach are significant, as optimizing chest compression parameters during CPR can improve the chances of successful resuscitation and reduce the risk of complications. The five key signals that are critical for effective CPR operations have been selected for experimentation: Compression, Blood Pressure, Velocity, Force, and Pmouth \citep{babbs05}. These signals provide essential insights into the physiological response during chest compressions and the overall effectiveness of CPR. Compression tracks the depth and rate of compressions, while Blood Pressure monitors circulatory efficacy. Velocity measures the speed of the chest recoil, ensuring optimal blood flow, and Force indicates the pressure applied during compressions to maintain perfusion. Pmouth reflects airway pressure, which is vital for ensuring proper ventilation. Together, these signals offer a comprehensive understanding of CPR performance, and a sample of these signals for a single patient is depicted in Figure \ref{fig: clean_data.}.

\subsubsection{Add Noise and artifacts to simulated data} \label{Sec:Noise}
To enhance the realism of the simulated CPR data generated by the Babbs model, several noise artifacts are incorporated. These additional noise components are strategically introduced to mimic the diverse sources of interference commonly encountered in real-world CPR signals. By augmenting the simulated data with realistic noise, a more faithful representation of the complexities and challenges inherent in CPR monitoring scenarios is aimed to be created. This approach ensures that the research experiments and algorithm developments are robustly tested against a spectrum of noise patterns, thus increasing the validity and applicability of the findings to real-world clinical settings. Through this meticulous process of noise augmentation, the gap between simulated and actual CPR data strives to be bridged, ultimately advancing the field of CPR monitoring and signal processing with greater fidelity and relevance.\\

\textbf{Gaussian Noise: }\\
In the context of CPR signal data, adding Gaussian noise is a crucial step in enhancing the realism and authenticity of simulated data for research purposes. Gaussian noise, also known as white noise, follows a normal distribution characterized by its mean $\mu$ and variance $\sigma^2$: \\
\begin{align}
Noise = \mu + \sigma \times \mathcal{N}(0, 1)
\end{align}\\
For this study, we tested specific parameter values: a mean of 0 and a standard deviation (or variance) of 1. These values were selected to simulate typical noise conditions encountered during CPR. To reflect more extreme conditions, such as a noisy ambulance environment, we increased the standard deviation to introduce greater variability and mimic higher levels of interference. Additionally, by setting a noise factor of 1.2 and a noise probability of 0.1, we controlled the intensity and likelihood of noise affecting the clean CPR signals. This controlled introduction of variability captures the stochastic nature of real-world noise during CPR monitoring, ensuring that the data more accurately reflects conditions encountered in clinical settings. \citep{gaussian00}.\\\\
%
% Additionally, the noise probability parameter determines the likelihood of noise being added to each element of the signal, further enhancing the real-life scenario of the simulated data. This process effectively replicates various sources of interference, such as electrical noise or sensor inaccuracies, which can impact the accuracy and reliability of CPR signal measurements. By adding Gaussian noise to the clean CPR data, the robustness of algorithms and methodologies can be evaluated under more realistic conditions \citep{gaussian02}. This facilitates advancements in CPR monitoring and decision-making in clinical settings, ensuring that our solutions are well-equipped to handle the complexities and challenges of real-world scenarios.
Additionally, the noise probability parameter controls the likelihood of noise being added to each element of the signal, further enhancing the realism of the simulated data. This approach replicates various sources of interference, such as electrical noise, sensor inaccuracies, or environmental disturbances. For example, real-world CPR monitoring scenarios often involve noise from moving equipment, patient movement during compressions, or chaotic settings like ambulances where vibrations and varying patient profiles can introduce significant interference. By incorporating Gaussian noise into the clean CPR data, the algorithms and methodologies can be rigorously evaluated under more practical conditions that closely mimic the unpredictable nature of clinical environments \citep{gaussian02}. This noise augmentation process facilitates the development of more robust CPR monitoring and decision-making systems, ensuring the proposed solutions can effectively address the challenges posed by real-world scenarios.\\

\textbf{Salt \& Paper Noise: }\\
In investigating CPR signal data, salt-and-pepper noise is introduced to simulate impulsive disturbances often encountered in real-world monitoring environments. This noise type mimics issues like sensor anomalies or transmission irregularities, which can affect the precision of CPR signal assessments. The chosen parameters for this noise, include salt probability = 0.0001, pepper probability = 0.0001, and salt and pepper values = 0.00001. These values reflect typical error margins in sensor readings and momentary transmission glitches observed during CPR monitoring. For instance, low probabilities such as 0.0001 were selected to represent rare but impactful occurrences, such as sudden malfunctions in medical equipment or brief disruptions in signal transmission, which are infrequent but can lead to significant distortions in the data. The salt and pepper values (0.00001) represent extremely low and high-intensity values, simulating hardware malfunctions that inject sharp spikes or drops into the data \citep{salt_00}. This methodology ensures the simulated data realistically mimics real-life noise scenarios, where occasional, sudden, and extreme disturbances can impair signal integrity during CPR operations. By simulating these conditions, our approach enables the robust evaluation of algorithms, ensuring they are equipped to handle the unpredictability of actual clinical environments \citep{salt_01}.\\

% In the investigation into CPR signal data, the introduction of salt and pepper noise is integrated using a specified function. This type of noise, characterized by sporadic occurrences of extreme intensity values, mimics the impulsive noise commonly encountered in practical monitoring environments. By employing the function with parameters such as salt probability = 0.0001, pepper probability = 0.0001, salt value = 0.00001, and pepper value = 0.00001, random instances of high-intensity "salt" signals and low-intensity "pepper" signals are injected into the original CPR signals \citep{salt_00}. This methodology aims to replicate various sources of interference, such as sensor anomalies or signal transmission irregularities, which may distort the precision of CPR signal assessments. The generation of salt and pepper noise involves the random selection of signals within the signal and substituting them with the specified salt and pepper values based on the provided probabilities. Through this approach, the simulated CPR data achieves greater fidelity to real-world monitoring scenarios \citep{salt_01}.\\

\textbf{Simulate Baseline Wander: }\\
The $Baseline~Wander$, modeled as a sinusoidal waveform, introduces subtle variations that capture the complexities inherent in CPR signal analysis, where $T$ represents the period of the sine wave (time for one complete cycle) and $t$ instantaneous time at which the function is evaluated. It is simulated using a function with an $amplitude$ parameter set at 0.02 to replicate the low-frequency variations often caused by physiological factors such as body movement or electrode displacement. These variations can distort the primary CPR signal, and the chosen amplitude reflects realistic levels of baseline wander observed in clinical settings. The 0.02 amplitude ensures that the simulation introduces sufficient variability without overwhelming the core signal, effectively emulating real-world noise sources like patient movement or sensor shifts during chest compressions in emergencies. This approach allows for a more accurate representation of the noise challenges encountered during CPR monitoring \citep{baseline_01}.

\begin{align}
Baseline~Wander = amplitude \times \sin\left(\frac{2\pi t}{T}\right)
\end{align}\\
This added realism enhances the robustness of our dataset, allowing for more comprehensive testing and refinement of algorithms. By incorporating these nuanced fluctuations, the dataset becomes better suited for exploring diverse research directions and improving algorithmic performance, ultimately advancing the reliability of CPR monitoring and analysis techniques.\\

\textbf{Muscle Interference: }\\
In our data augmentation process for CPR signal data, $Muscle~Inter\!f\!erence$ has resembled using a function with a specified amplitude parameter of 0.05. Muscle interference represents the unpredictable noise that arises from involuntary muscle contractions, movement artifacts, or electrical noise from surrounding muscles and it masks the features of the underlying signal. By adding muscle interference with this parameter, realistic variations in the CPR signals are simulated, mimicking the effects of muscle-related artifacts that are observed in real-world monitoring scenarios. The simulated interference is generated as Gaussian noise with a mean of 0 and a standard deviation corresponding to the specified amplitude, which enriches the CPR signals by introducing nuanced noise patterns that represent the complexities of CPR signal \citep{muscle_00}.

\begin{align}
Muscle~Inter\!f\!erence = amplitude \times \mathcal{N}(0, 1)
\end{align}\\
This augmentation will enhance the resilience of CPR signals and facilitate more comprehensive research exploration and algorithm development, which will help to improve the accuracy and reliability of CPR signal processing algorithms.\\

\textbf{Simulate Sudden Amplitude Changes: }\\
Moreover, a unique approach is implemented by applying the simulated sudden amplitude changes function. This function is executed with several parameters such as num changes = 500, max change duration = 10, and change factor = 0.005. At the same time, a sudden variation in signal intensity is injected which resembles the abrupt fluctuations often encountered in real-world scenarios. These changes reflect the artifacts that arise from equipment glitches, physiological shifts, or external disruptions, introducing an element of unpredictability into the signals. By employing this function with carefully chosen parameters, numerous instances of sudden amplitude changes are simulated, each limited to a maximum duration of 10 data points. These parameters were meticulously selected to mimic the dynamic nature of sudden amplitude shifts observed during CPR monitoring scenarios. Through these amplitude changes, our CPR signals gain an added layer of realism and encourage deeper research exploration \citep{amplitude_00}.\\

{\textbf{Simulated Compression Depth Variations: }\\
In the ongoing efforts to enrich CPR signal data, the simulated compression depth variations function is applied by applying parameters such as num variations = 500, max variation duration = 20, and variation factor = 0.8. This custom function is designed to replicate fluctuations in compression depth which is a common occurrence in medical signals due to changes in applied pressure or physiological dynamics. By strategically adjusting these parameters, a wide range of compression depth variations is introduced into our signals. These parameters were deliberately chosen to mirror the unpredictable nature of compression depth changes observed in real-world CPR monitoring scenarios and provide a faithful representation of dynamic signal characteristics. This augmentation strategy enhances the adaptability of CPR signal processing algorithms which enables them to effectively navigate fluctuations in compression depth with accuracy and precision \citep{compress_00}.\\

{\textbf{Simulate Multiple Dropouts: }\\
Finally, the multiple dropouts function has been applied by configuring with parameters such as num dropouts = 500 and max dropout duration = 10. This function replicates multiple instances of dropouts in the signal and mimics periods where data points are missing or corrupted. By implementing these parameters, numerous dropout instances are introduced into the input data frame where the signal values are substituted with NaN (Not a Number) to indicate the absence of data. Missing the signal values is very common in real-life scenarios in the case of CPR and is very crucial, whereas the integration of simulated dropouts strengthens the adaptability of CPR signal processing algorithms in handling missing or corrupted data points \citep{dropout_00}.\\

We applied all of these artifacts to the signals, effectively transforming them into realistic representations akin to the simulated signals generated by the Babbs model. By incorporating Gaussian noise, salt and pepper noise, baseline wander, muscle interference, sudden amplitude changes, compression depth variations, and multiple dropouts, the signals mimic the complexities and challenges encountered in real-life CPR monitoring scenarios. Examples of noisy signals from a single patient are presented in Figure \ref{fig:Noisy_signal}. As a result, each artifact introduces its unique distortions and irregularities, enriching the dataset with diverse noise patterns and signal variations. This comprehensive augmentation process not only enhances the fidelity of our dataset but also prepares it to handle real-world signal processing challenges.

\begin{figure}[htbp]
    \begin{center}
    \centering
    \includegraphics[width=1\textwidth,height=.28\textwidth]{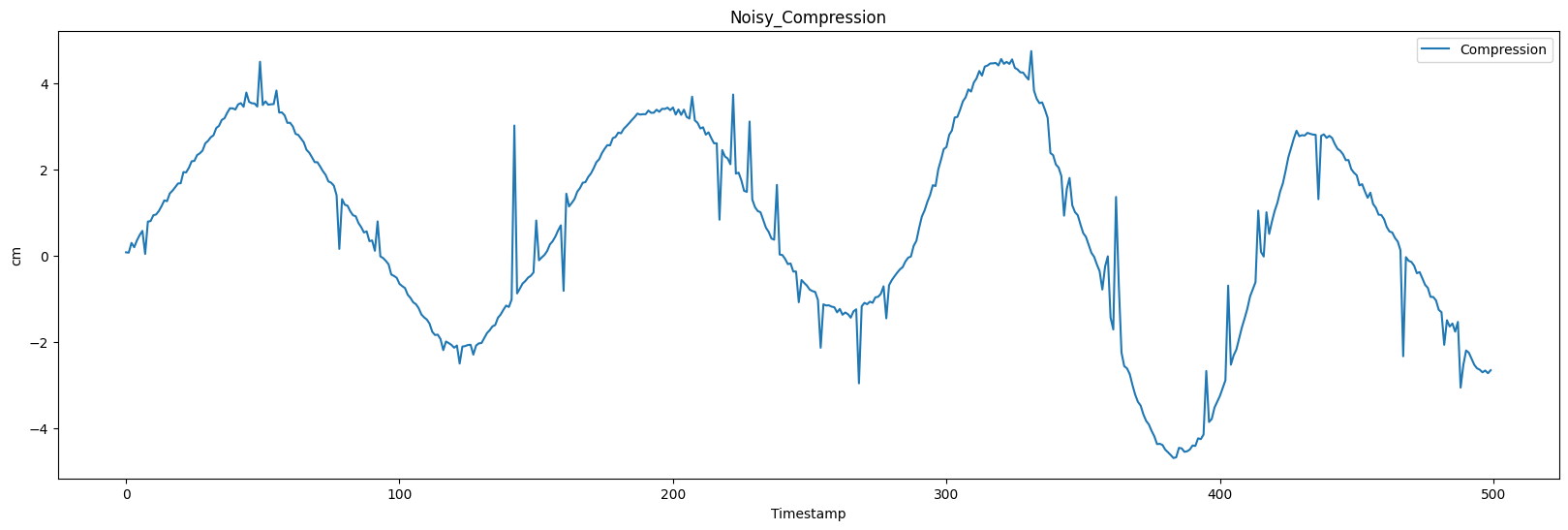}
    \includegraphics[width=1\textwidth,height=.28\textwidth]{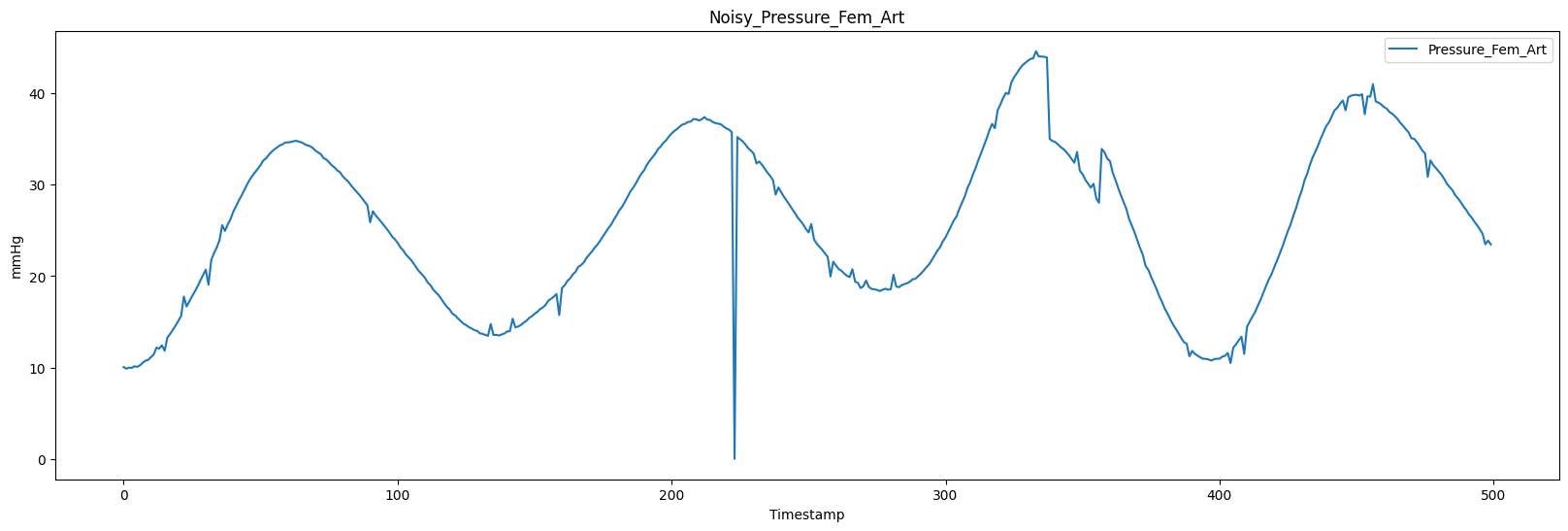}
    \includegraphics[width=1\textwidth,height=.28\textwidth]{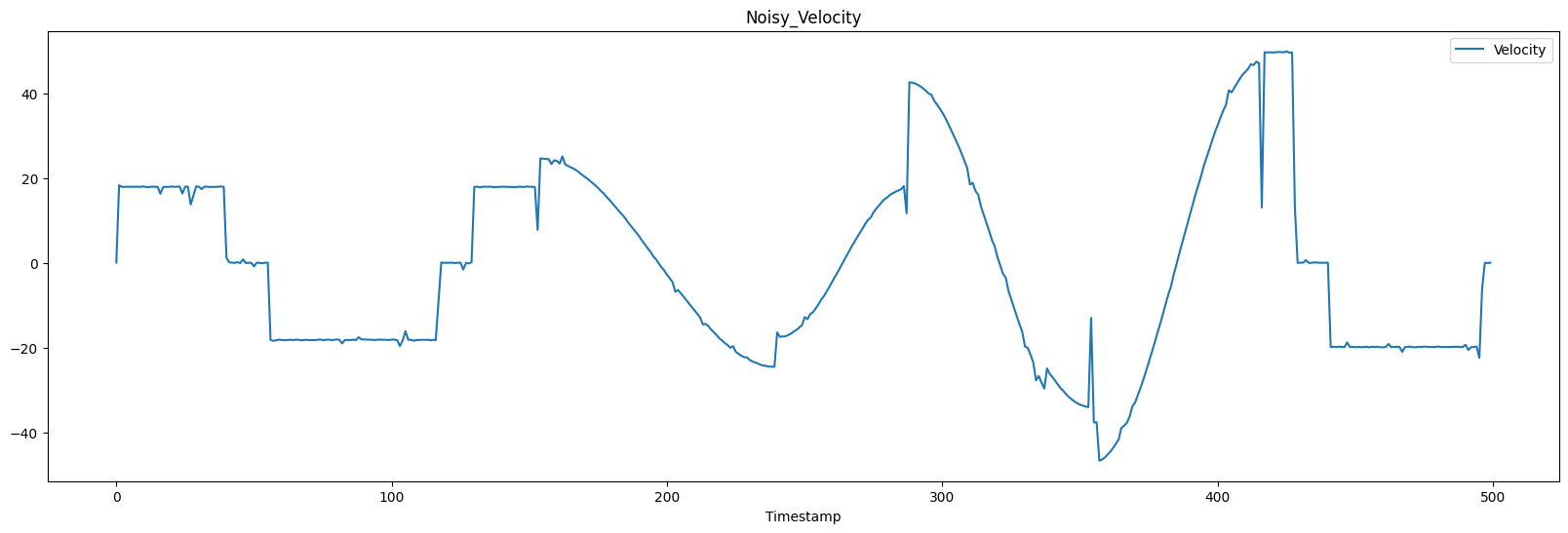}
    \includegraphics[width=1\textwidth,height=.28\textwidth]{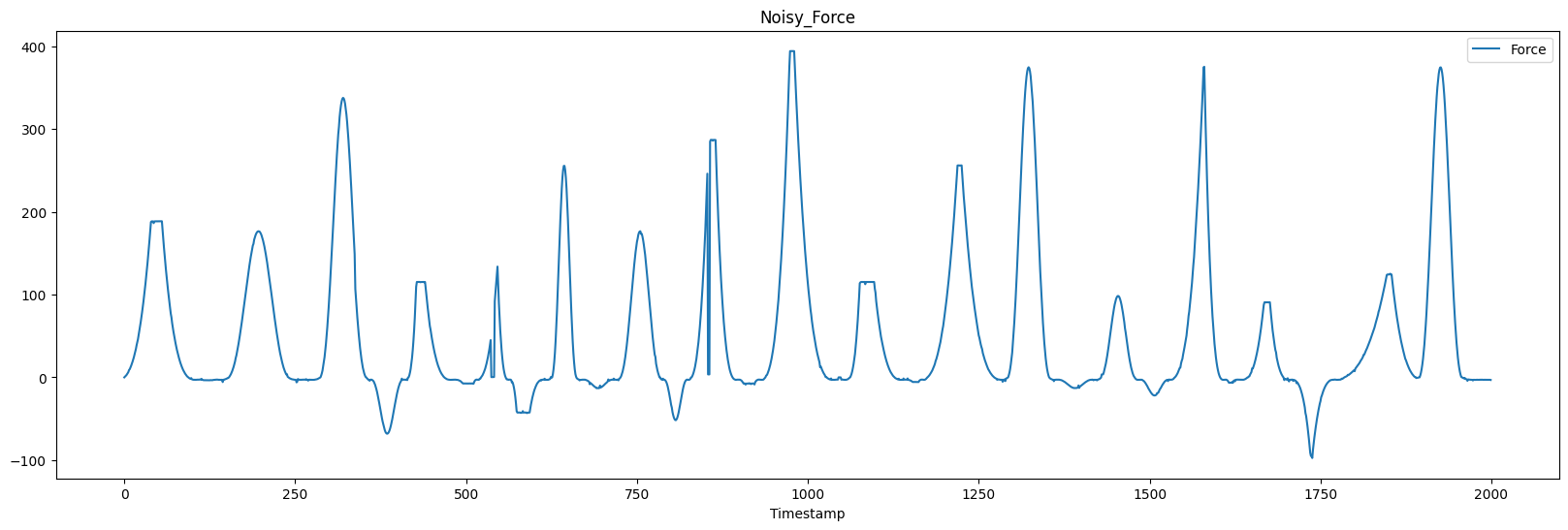}
    \includegraphics[width=1\textwidth,height=.28\textwidth]{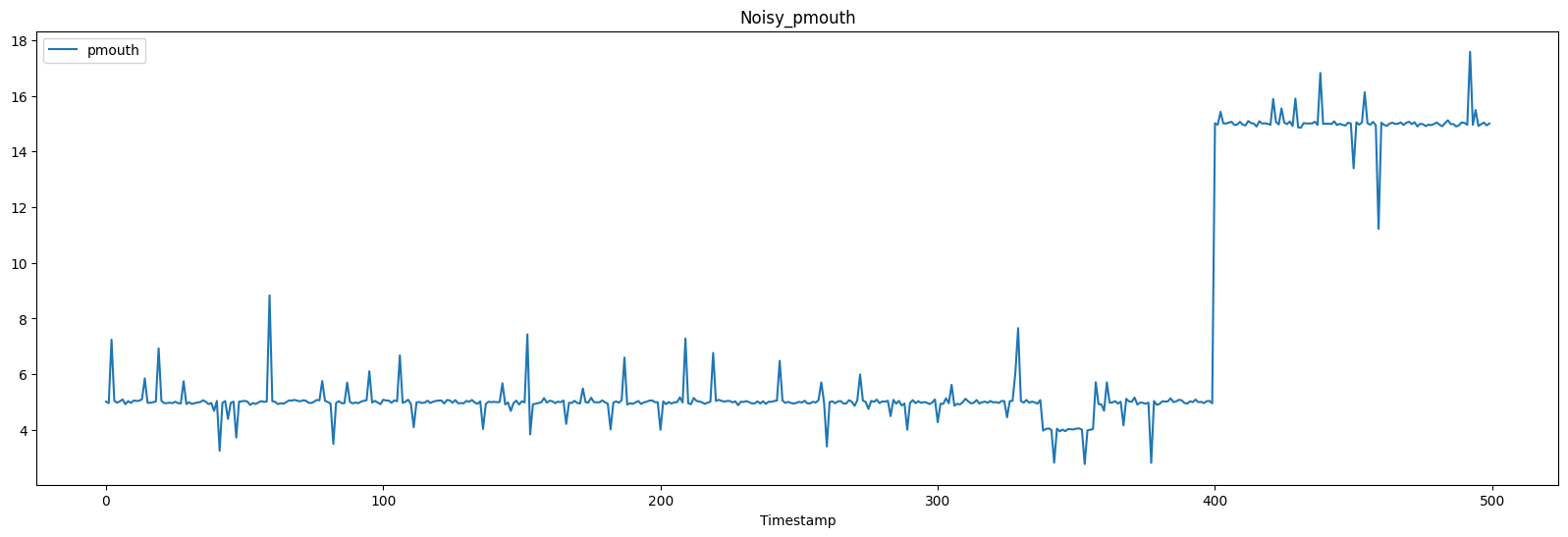}
    \caption{Noisy CPR - Compression, Pressure, velocity, Force, Pmouth signals.}
    \label{fig:Noisy_signal}
    \vspace{-5pt} % reduce the space after the picture by 10pt
    \end{center}
\end{figure}

\subsection{Experiment Set-up and training} \label{Sec:Model_training}

During the model training, we employed a set of Python libraries, including TensorFlow, Scikit-learn, Matplotlib, NumPy, and Pandas, among others, to facilitate various aspects of the training process. We trained our model in an unsupervised manner with noisy CPR signals of only three(3) different patient data and split the data into train(80\%) and validation(20\%) sets for training purposes. After several experiments with the model's hyperparameters, our model was trained using a batch size of 64 instances per iteration which provided a balance between computational efficiency and model convergence. To mitigate the risk of overfitting, a common challenge in deep learning tasks, we incorporated a callback function with a patience parameter set to 3, allowing the model to train until no improvement in validation loss was observed for three consecutive epochs \citep{overfit_00}. This approach helped to prevent the model from memorizing noise in the training data and encouraged it to learn widespread patterns. For the optimization strategy, we opted for the popular Adam optimizer, known for its effectiveness in training deep neural networks by adapting the learning rates for individual parameters \citep{adam_00}. Additionally, we employed the Mean Absolute Error (MAE) loss metric, a robust choice for regression tasks that provided a clear and interpretable measure of the model's performance in minimizing prediction errors \citep{MSE_00}. By carefully selecting these training parameters and strategies, we aimed to optimize the training process and maximize the model's ability to learn meaningful representations from the data, ultimately enhancing its denoising capabilities for biomedical signal processing applications.

\section{Results and Analysis} \label{Sec:result}

Our methodology was trained for nine epochs, as detailed in Section \ref{Sec:Model_training}, striking a balance between allowing the model to learn from the data and minimizing computational overhead. The training and validation loss, illustrated in Figure \ref{fig:loss_validation}, indicate effective learning: the training loss (blue line) starts at approximately 0.14 and steadily decreases to around 0.02 by the ninth epoch, while the validation loss (orange line) begins at about 0.16 and declines to roughly 0.03. Both loss curves demonstrate a downward trend that converges towards the end, suggesting the model is not overfitting, given the close proximity of the training and validation losses. Upon completing the training, we applied our model in inference mode to the noisy signals of a new patient, enabling a visual assessment of its performance across five key CPR signals. The visualizations in Figure \ref{fig:final_result} compare the noisy signals with their denoised counterparts, clearly showing that the proposed methodology effectively reduces noise. The smoother lines of the denoised signals closely follow the trends of the original noisy signals across various contexts, including pressure, compression, activity, voice, velocity, and force, thereby enhancing clarity and interpretability. Overall, the results affirm the robustness and effectiveness of our approach in denoising signals across diverse applications, demonstrating that our model has significantly reduced noise and improved signal quality.

\begin{figure}[htbp]
    \begin{center}
    %\advance\leftskip 1.8 cm
    \centering
    \includegraphics[width=.65\textwidth,height=.5\textwidth]{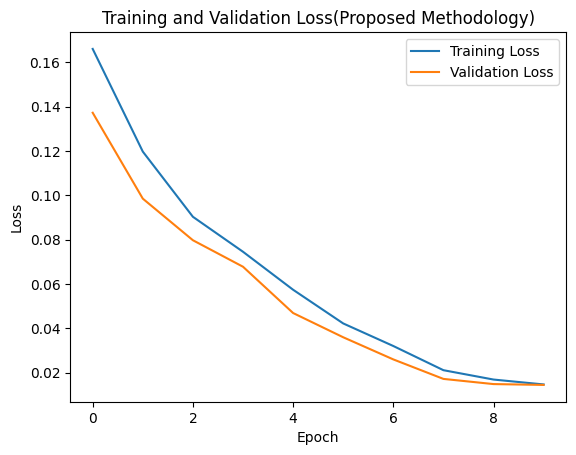}
    \caption{Training and validation loss of model with Multi-modal approach.}
    \label{fig:loss_validation}
    \vspace{-10pt} % reduce the space after the picture by 10pt
    \end{center}
\end{figure}

\begin{figure}[htbp]
    \begin{center}
    \centering
    \includegraphics[width=.9\textwidth,height=1.4\textwidth]{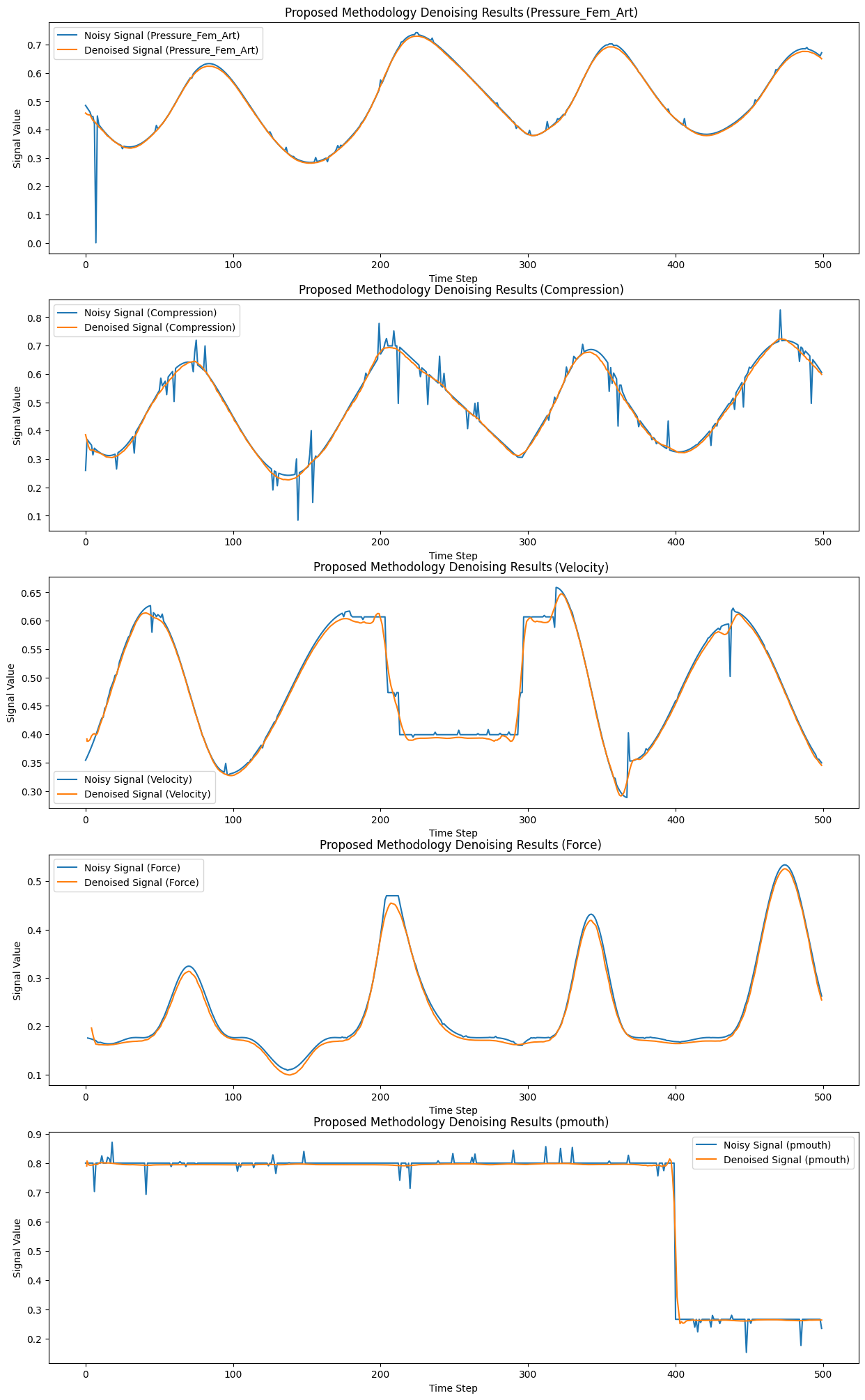}
    \caption{Denoised signals with the proposed methodology.}
    \label{fig:final_result}
    \vspace{-5pt} % reduce the space after the picture by 10pt
    \end{center}
\end{figure}

\subsection{Comparision with existing methods}

To comprehensively evaluate the effectiveness of our proposed framework, we conducted a comparative analysis against existing ML and filtering methodologies. Given that one of the primary objectives of our methodology is to achieve robust performance in an unsupervised manner, the comparisons were conducted under unsupervised conditions. For the filtering approach, we employed the enhanced adaptive filter proposed by \cite{related_01}, which is an advanced method specifically designed for CPR signal processing. Since filtering techniques by default operate without labeled data, we applied the adaptive filter as outlined in \cite{related_01}, with fine-tuning of parameters to better accommodate the different CPR signals in our study. For the existing ML-based method, we considered the approach proposed by \cite{related_07}, which is currently the only dedicated ML method for CPR signal denoising and which was originally designed as a supervised approach. To enable a fair comparison with our unsupervised framework, we adapted their method to operate in an unsupervised manner to the best extent possible. Both the filter-based and adapted ML methods were trained and tuned on the same CPR signals used for our proposed approach. We subsequently tested these methods on the same new patient signals, enabling a direct comparison of their denoising performance against our framework. Figure \ref{fig:comparison_results} illustrates the resulting denoised signals: the red line represents the noisy signal, the green line depicts the denoised signals produced by our proposed method, the yellow line shows the denoising results obtained from applying an existing ML method in an unsupervised approach, and the blue line displays the results achieved using a filtering technique. The visualizations demonstrate that our proposed method effectively captures complex signal patterns while successfully removing artifacts, resulting in a cleaner, more robust signal. Although the filter-based approach performed reasonably, it struggled to identify intricate patterns within the signals, leading to poor denoising in certain areas. On the other hand, the existing ML approach without labeled data, showed poor performance. It failed to capture even simple patterns within the signal, resulting in a highly distorted output. The filter-based method is performed at a certain level; however, considering human life and death matters like CPR, it is necessary to ensure a high level of precision at any cost, which is possible by the proposed methodology. So, upon analyzing these visualizations, it is evident that our methodology consistently outperforms the existing approaches especially without labeled data, offering superior denoising quality. This advantage is especially critical in biomedical applications like CPR, where signal clarity is essential. The effectiveness of our approach is further validated through the visual comparisons and the quantitative metrics applied in this study, underscoring the value of our framework in enhancing CPR signal processing.

\begin{figure}[htbp]
    \begin{center}
    \centering
    \includegraphics[width=.8\textwidth,height=1.4\textwidth]{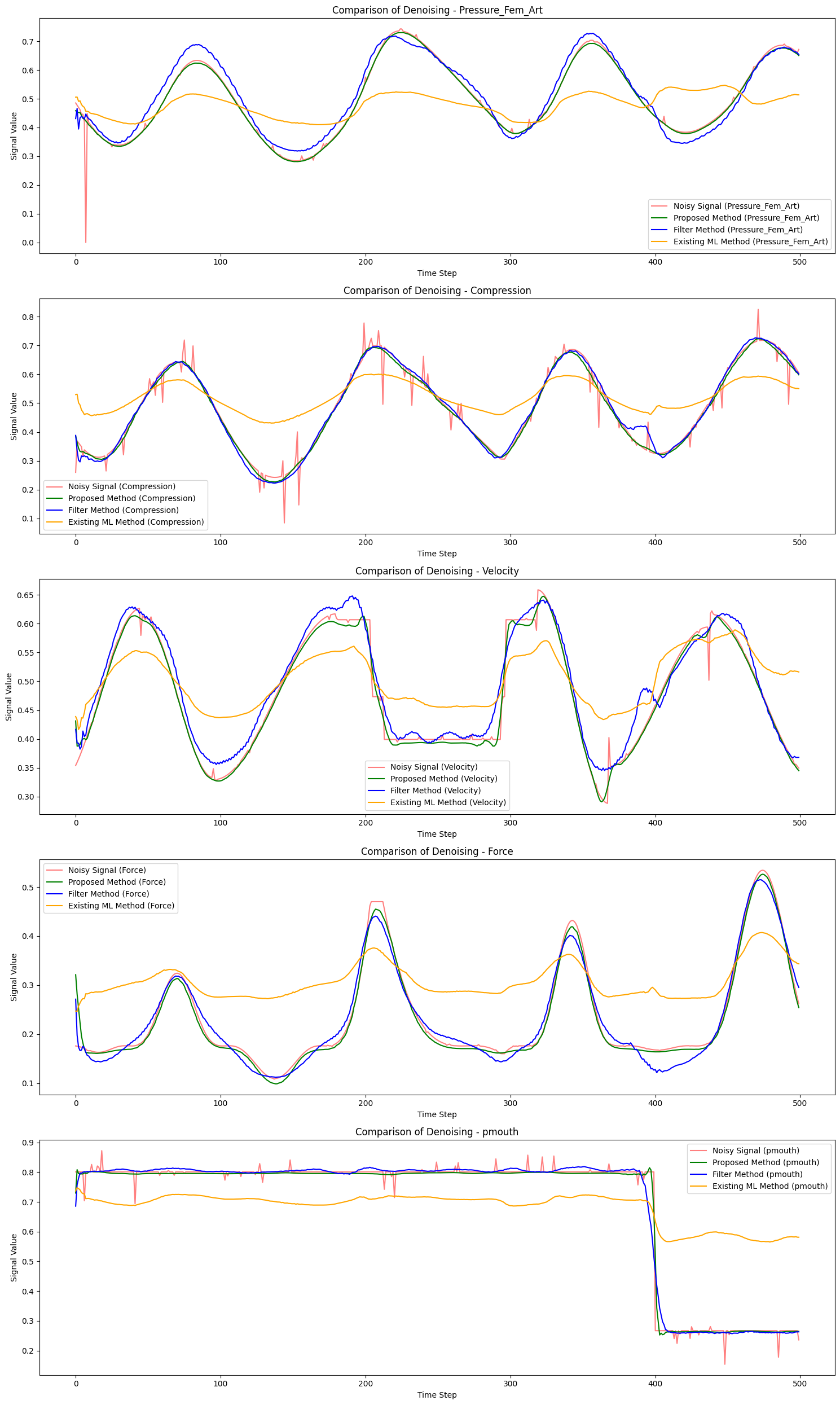}
    \caption{Comparison of CPR signal denoising performance between the proposed, existing ML, and filter methods in an unsupervised context.}
    \label{fig:comparison_results}
    \vspace{-5pt} % reduce the space after the picture by 10pt
    \end{center}
\end{figure}

\subsection{Performance Analysis}
\textbf{Signal-to-Noise Ratio:}
Signal-to-Noise Ratio ($SNR$) is a fundamental metric used in signal processing to quantify the quality of a signal by comparing the strength of the signal to the level of noise. $SNR$ is defined as the ratio of the power of the signal to the power of the noise. In mathematical terms, $SNR$ is often expressed in decibels ($dB$) using a logarithmic scale, which allows for easy interpretation of the ratio. A higher $SNR$ value indicates that the signal is stronger relative to the noise, which generally leads to better signal quality and improved performance in various applications. $SNR$ plays a crucial role in determining the reliability and accuracy of signal-processing algorithms, making it an essential concept in multiple fields. It provides a quantitative measure of signal quality relative to background noise \citep{SNR_01}. Mathematically, $SNR$ is defined as the ratio of the power of the signal ($S$) to the power of the noise ($N$). This ratio is often expressed in decibels ($dB$) using a logarithmic scale:

\begin{align}
SNR_{dB} = 10 \cdot \log_{10} \left( \frac{S}{N} \right)
\end{align}\\
\textbf{Peak Signal-to-Noise Ratio:}
Moreover, Peak Signal-to-Noise Ratio ($PSNR$) is a specific form of SNR popularly used in image processing to evaluate the image noise. Its underlying concept is also applied to other types of data like signals. $PSNR$ measures the ratio between the maximum possible power of a signal $\mathbf{M}$ and the power of the noise introduced by compression or other forms of distortion. $PSNR$ is expressed in decibels ($dB$), allowing for easy comparison of signal quality across different techniques. A higher $PSNR$ value indicates that the denoised signal is closer to the original, noise-free signal, suggesting better signal quality with less distortion or loss of information \citep{PSNR_00}. Unlike $SNR$, $PSNR$ is often calculated based on continuous signals and expressed in decibels ($dB$), which is calculated using the mean squared error ($MSE$) between the original noise-free signal ($S$) and the distorted signal ($S$):

\begin{align}
PSNR_{\text{dB}} = 10 \cdot \log_{10} \left( \frac{\mathbf{M}^2}{MSE} \right)
\end{align}\\
%
% We considered both SNR and PSNR to analyze our proposed methodology with the same autoencoder architecture as the conventional method. The bar graph in Figure \ref{fig:SNR_and_PSNR} presents a comparison between the conventional and proposed methods concerning SNR and PSNR metrics for evaluating signal denoising quality. Upon examination, the Proposed method exhibits significant superiority over the Conventional method in both SNR and PSNR values. Specifically, the Proposed method achieves a markedly higher SNR value of 27.73 compared to 23.06 for the Conventional method, indicating a stronger signal relative to background noise. Additionally, the Proposed method demonstrates superior signal fidelity with a PSNR value of 33.89, substantially surpassing the Conventional method's PSNR of 28.71. This disparity underscores the efficacy of the proposed method in noise reduction and signal preservation, offering enhanced reliability and performance for applications requiring high-quality signals. Overall, the results advocate for the adoption of the Proposed method as a promising solution for improving signal integrity and quality during CPR.\\
%
Both $SNR$ and $PSNR$ are used to analyze the effectiveness of the proposed methodology and compare it with the existing ML and filter methods. The bar graph in Figure \ref{fig:SNR_and_PSNR} compares the proposed, filter and existing ML methods without labeled data concerning $SNR$ and $PSNR$ metrics for evaluating signal denoising quality. The proposed method demonstrates substantial improvements over both of the existing methods in terms of $SNR$ and $PSNR$ metrics. Specifically, our proposed method achieves a significantly higher $SNR$ of 27.73, outperforming the filter and the existing ML approaches, which yield $SNR$ values of 21.7 and 12.81, respectively. This highlights our method's enhanced capability to preserve the true signal while effectively minimizing background noise. Furthermore, the proposed approach exhibits superior signal fidelity, as reflected by a $PSNR$ of 33.89, which far exceeds the $PSNR$ values of 25.32 and 16.53 obtained by the filter method and the existing ML technique, respectively. This difference highlights the effectiveness of the proposed method in noise reduction and signal preservation, offering enhanced reliability and performance for applications requiring high-quality signals. Overall, the results support the adoption of the proposed method as a promising solution for improving signal integrity and quality during CPR.\\

\begin{figure}[htbp]
    \begin{center}
    %\advance\leftskip 1.8 cm
    \centering
    \includegraphics[width=.65\textwidth,height=.49\textwidth]{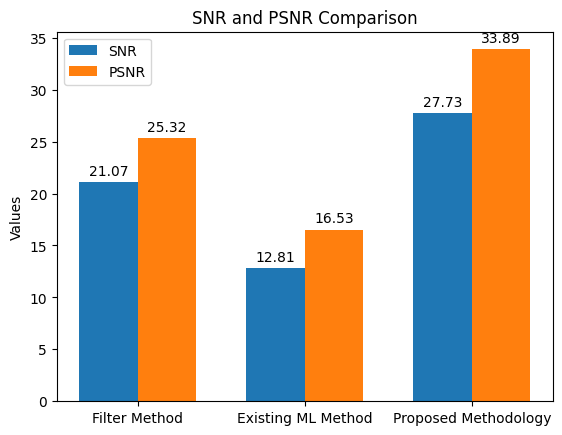}
    \caption{SNR and PSNR comparison between the proposed, existing ML and filter methods in an unsupervised context.}
    \label{fig:SNR_and_PSNR}
    \vspace{-10pt} % reduce the space after the picture by 10pt
    \end{center}
\end{figure}

\begin{figure}[htbp]
    \centering
    \includegraphics[width=0.48\textwidth]{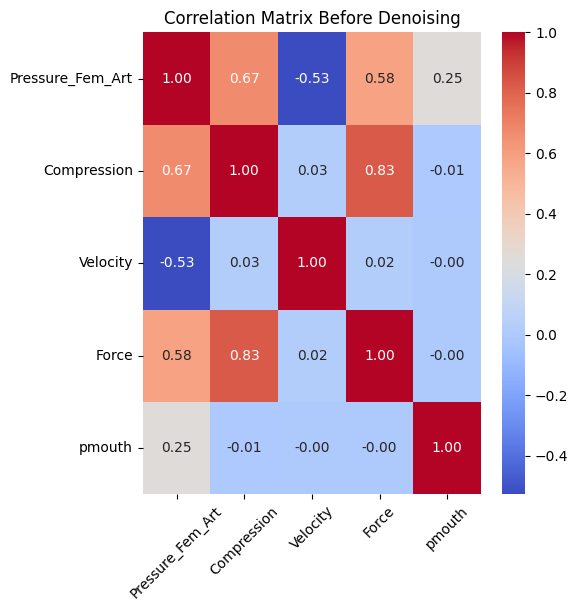}
    \hfill
    \includegraphics[width=0.48\textwidth]{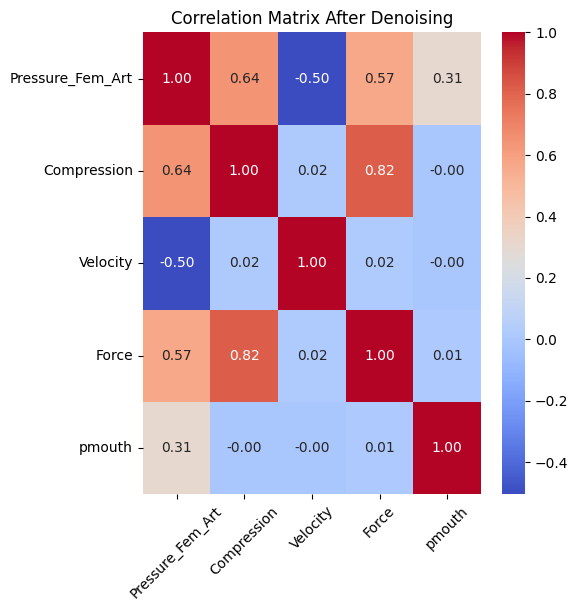}
    \caption{Correlation matrix before denoising and after denoising with proposed methodology and framework.}
    \label{fig:Correlation_matrix}
    \vspace{-10pt}
\end{figure}

Another critical aspect of the framework and methodology is the preservation of data correlation during the denoising process. Maintaining the correlation within the data is essential for subsequent machine learning applications in downstream tasks, such as blood pressure measurement, and the optimization of compression and decompression processes, after the removal of signal artifacts. In our study, we carefully evaluated the impact of our denoising framework and methodology on the correlation matrix of our dataset. First, we calculated and visualized the data correlation matrix before and after applying our denoising techniques \citep{Correlation_00}. We found remarkably similar correlational values in both instances, as shown in Figure \ref{fig:Correlation_matrix}. This observation indicated that our framework effectively preserved the intrinsic relationships among the variables in the dataset. To further examine the correlation, we computed and compared the correlation coefficients between the two correlation matrices, as shown in Figure \ref{fig:Correlation_comparison}. The correlation coefficient measures the similarity between two correlation matrices, with values ranging from -1 to 1; higher values denote better similarity \citep{Correlation_01}. We obtained a correlation coefficient value of 0.9993, signifying an exceptionally high degree of similarity between the two matrices. This finding underscores the robustness and efficacy of our denoising methodology in maintaining the underlying correlations within the dataset while effectively denoising signals. Thus, the proposed framework and methodology validate its potential to be utilized for enhanced denoising reliability and maintaining data quality in various biomedical applications, including CPR signals \citep{Co_co_00}.\\

\begin{figure}[ht]
    \begin{center}
    %\advance\leftskip 1.8 cm
    \centering
    \includegraphics[width=.58\textwidth,height=.44\textwidth]{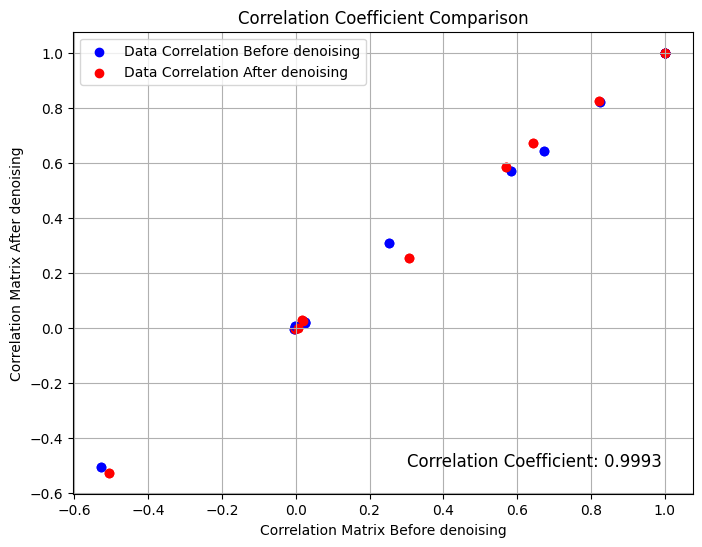}
    \caption{Data correlation coefficient in before and after denoising through our proposed methods.}
    \label{fig:Correlation_comparison}
    %\vspace{-20pt} % reduce the space after the picture by 10pt
    \end{center}
\end{figure}

To recapitulate, the proposed methodology and framework provide significant advantages for biomedical signal denoising. Firstly, the multi-modality approach allows for the simultaneous processing of multiple signals without the need for labeled data, leveraging correlations to enhance denoising performance and improve signal quality, which is particularly beneficial for CPR automation using ML techniques. Secondly, individual denoising of each signal targets their unique characteristics and noise profiles, leading to more effective noise removal and overall enhanced performance. The framework's flexibility accommodates diverse signal modalities and noise sources, allowing experimentation with various ML models tailored to specific signal characteristics. Additionally, its scalability efficiently handles complex denoising tasks involving large data volumes. By denoising each signal separately before combining them, the framework ensures robustness against noise fluctuations, ultimately facilitating more accurate clinical interpretation and decision-making without relying solely on labeled data or traditional filtering methods.

\section{{Discussion} \label{future_work}}

One of the key objectives is to validate the CPR data generated by the Babbs model using real patient data, once access to real-life CPR data from medical authorities is obtained through our ongoing collaboration with cardiologists at the Montreal Heart Institute. This will allow us to verify the data distribution and assess the similarity between the simulated and real-world data. By validating the synthesized data against actual clinical conditions, we aim to ensure its accuracy and reliability. After this validation process, our goal is to make the synthesized data publicly available for research purposes, promoting wider validation and further exploration of the algorithm's potential in improving CPR outcomes.\\

Moreover, the successful development of the multi-modal ML framework for denoising biomedical signals opens up several avenues for future research and application. In the forthcoming stages of this research, the model exploration, particularly focusing on integrating cutting-edge hybrid architectures as the base model of autoencoder in our framework can bring significant improvement. While different signals may benefit from specific architectural features, applying the same architecture-based autoencoder for each modality or signal is not essential in extensive application. However, it’s essential to recognize that certain signals may inherently benefit from alternative architectures-based autoencoders, which excel in capturing temporal dependencies or focusing on specific regions of interest for each signal. For instance, combining transformer and CNN models, such as Convit, Conformer, and related innovations offer compelling advantages of multiple algorithms through a single architecture. For example, these models adapt the advantages of both transformer and CNN; however, transformers are adept at capturing long-term dependencies inherent in sequential data, while CNN is proven to capture local features and patterns \citep{future__00}. By assessing the performance of such hybrid models within the framework of our denoising methodology, it would be possible to unlock new avenues for enhancing the robustness of the denoising process across biomedical signal modalities. Besides, other biomedical signal processing specialized transformer-based models, like, TransHFO, CRT-Net, CAT, etc can be experimented with in their denoising performance as the base of the autoencoder in the framework \citep{ISLAM2024122666}. Furthermore, we can utilize the knowledge from the ML framework for denoising to make progress in downstream tasks by applying techniques like transfer learning or few-shot learning, which can be used in the learned representations to adapt easily to new tasks or areas, increasing the proposed framework's effectiveness in real-world situations. In addition, another aim in the future involves validating the simulated data generated by the Babbs model against real-life CPR data in collaboration with healthcare centers and hospitals. Once validated, publishing this CPR data for future research purposes will provide a valuable resource, enabling broader exploration and advancements in this critical area.\\

\begin{figure}[ht]
    \begin{center}
    %\advance\leftskip 1.8 cm
    \centering
    \includegraphics[width=1\textwidth,height=.3\textwidth]{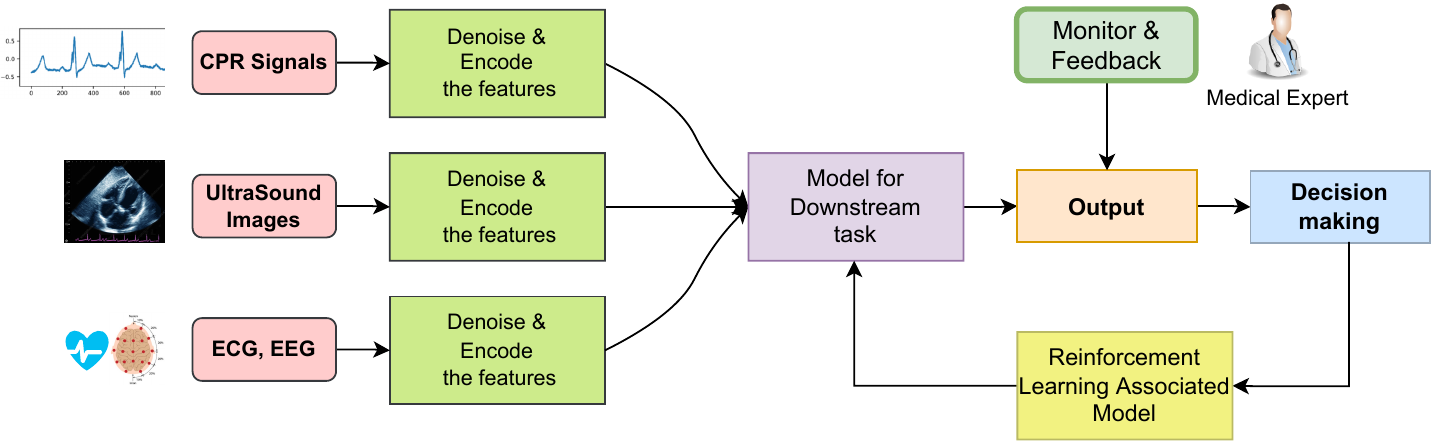}
    \caption{A multi-modal ML concept for decision marking in medical emergency.}
    \label{fig:future_Work}
    \vspace{-5pt} % reduce the space after the picture by 10pt
    \end{center}
\end{figure}

Another promising direction is the extension of the multi-model ML approach to other types of biomedical data beyond CPR signals only. By adapting the framework to accommodate different types of medical modalities and characteristics, researchers can explore its applicability in diverse medical contexts, such as ECG, electromyography (EMG), electroencephalography (EEG), ultrasound images, and so on along with CPR signals, as depicted in Figure \ref{fig:future_Work}. Initially, each type of data can be denoised and its features encoded into lower dimensions using appropriate denoising ML models according to the focus and merit of the specific data. Then, the outputs from all modalities can be trained with a model for downstream tasks to enable appropriate decision-making. Before finalizing decisions, the outputs from the multi-modality model should be monitored and reviewed by medical experts. Based on feedback from clinical experts, the output can be refined before making final decisions. To further enhance and automate the entire process, reinforcement learning (RL) can be deployed to learn decision-making policies. By doing so, the RL agent would refine its strategies through feedback loops, optimizing the decision-making process and providing actionable insights. The predictive model processing data from various modalities would then integrate this RL-based policy, ensuring that decisions are continually improved and adjusted based on real-world inputs and outcomes. Through this concept, better, more precise, and well-informed decision-making would be possible by utilizing more information from different modalities about a patient in emergency medical settings.\\

% Additionally, further investigation into the interpretability of the learned representations and the integration of domain knowledge could significantly enhance the fairness and trustworthiness of the denoising process. This would make the system more suitable for clinical decision-making and diagnosis, as medical professionals would have greater confidence in its outputs. Moreover, incorporating principles of explainable AI (XAI) would further promote transparency by providing insights into how the model arrives at its conclusions, ultimately fostering trust and improving the adoption of AI-driven methods in healthcare. Finally, the proposed framework's scalability and generalizability should be explored through larger-scale studies and real-world applications, with collaborations from healthcare institutions and industry partners facilitating its deployment and evaluation in clinical settings while allowing for valuable feedback from medical professionals and patients.\\

Another key aspect is supporting effective explanations
of our output to our users. Valued starting points that fit well with
our current design and show promise are suggested in the current literature. \cite{future_HOLZINGER} indicate that fusing information from different modalities will be a valuable pathway for explaining AI-driven decision-making to medical professionals. They argue that the data collection can be enhanced by the modality that led to that data for a deeper analysis of the rationale for the AI system to suggest decision-making paths to diverse health experts. \cite{future_gu} point out that despite the black-box nature of CNNs (an important element of our proposed approach), generating the kind of explanations that health professionals may need can be achieved using attentional mechanisms to highlight why certain decisions are being proposed for the medical concern. They add that focusing on visual displays of key elements of the input data will be most important. A user study conducted by \cite{future_bienefeld} also showed the importance of considering the mental models of users when deciding both what information to display and what options to support within the visual display. Insights into how clinicians searched for deeper insights into the decision-making within the interface were very valuable to learn, towards progressive designs of their dashboard. To maintain effective interpretability of AI system results, \cite{future_Sendak} also points out the importance of ongoing feedback loops; continued endorsement from medical professionals and thus deployment are then possible. Their study concerned the medical condition of sepsis and the interaction with Emergency Medicine departments (AI systems that would be running and would be reacted to in real-time, as is the case with our study of CPR). \cite{future_Hui} in their extensive survey of explanation of AI systems for healthcare
singled out as a key concern the need to explain biosignal abnormalities towards clinical data interpretation; this suggests that our particular focus in deciding how to handle CPR in clinical settings using AI-based solutions is especially well chosen. The importance of denoising when working towards effective AI systems for healthcare decisions that include visual images is a point emphasized by \cite{future_Dhar} which also endorses our effort to develop an effective approach for this key consideration.

\section{Conclusion} \label{conclusion}

In conclusion, the significance of denoising biomedical signals, particularly in crucial life-saving interventions like CPR, cannot be overstated. because the accuracy and reliability of CPR data can be a matter of life and death. Despite advancements in technology, traditional methods of signal denoising using filters often fall short of effectively removing noise while preserving the integrity of underlying signals. However, a high level of precision is crucial considering CPR as a matter of human life and death. In this context, the capability of ML to capture the complex underlying patterns is proven in the biomedical domain. Though a few dedicated ML methods for denoising CPR signals can be notified in the domain, those are supervised approaches. However, the unavailability and difficulty of getting the labeled clean signals corresponding to the noisy signal hinder the existing ML method's potential in real-life contexts. To solve this problem, a dedicated unsupervised ML method for CPR signals is required in the domain, which can process multiple CPR signals concurrently without labeled data. The proposed multi-modal machine learning framework addresses this challenge by offering a solution to enhance the quality and utility of vital physiological measurements in an unsupervised manner, ensuring that noise reduction is achieved without compromising signal integrity. Notably, the methodology excels in preserving data correlation, which is crucial for downstream tasks. The contributions of the proposed methodology can be extended beyond CPR signal denoising, where the methodology can be effective in handling various types of biomedical signals and data, paving the way for broader machine-learning applications in healthcare and medical research.

\bibliographystyle{model5-names}
{\footnotesize
\bibliography{iclr2021_conference.bib}}

\clearpage

\end{document}